\documentclass[preprint,12pt,authoryear]{elsarticle}

\usepackage{amssymb}
\usepackage{amsmath}

\usepackage{caption}    
\usepackage{subcaption} 
\usepackage{graphicx}   
\usepackage{xfrac}
\usepackage{lmodern}
\usepackage{algpseudocode,algorithm,algorithmicx}
\usepackage{hyperref}
\usepackage{multirow}
\usepackage{multicol}
\usepackage{tikz}
\usetikzlibrary{calc}
\usetikzlibrary{er,positioning,bayesnet}
\tikzset{
    old inner xsep/.estore in=\oldinnerxsep,
    old inner ysep/.estore in=\oldinnerysep,
    double circle/.style 2 args={
        circle,
        old inner xsep=\pgfkeysvalueof{/pgf/inner xsep},
        old inner ysep=\pgfkeysvalueof{/pgf/inner ysep},
        /pgf/inner xsep=\oldinnerxsep+#1,
        /pgf/inner ysep=\oldinnerysep+#1,
        alias=sourcenode,
        append after command={
        let     \p1 = (sourcenode.center),
                \p2 = (sourcenode.east),
                \n1 = {\x2-\x1-#1-0.5*\pgflinewidth}
        in
            node [inner sep=0pt, draw, circle, minimum width=2*\n1,at=(\p1),#2] {}
        }
    },
    double circle/.default={2pt}{black}
}

\begin{document}

\begin{frontmatter}


\author[label1]{Dimitris G. Georgiadis\corref{cor1}}
\ead{dg.georgiadis@tum.de}
\cortext[cor1]{Corresponding author}
\affiliation[label1]{organization={Engineering Risk Analysis Group, Technical University of Munich},
            addressline={Theresienstrasse 90}, 
            city={Munich},
            postcode={80333}, 
            country={Germany}} 

\author[label2]{Manolis S. Samuelides}
\affiliation[label2]{organization={Division of Marine Structures, School of Naval Architecture and Marine Engineering, National Technical University of Athens},
            addressline={Iroon Polytechniou 9}, 
            city={Zografou},
            postcode={15772}, 
            country={Greece}}

\author[label3]{Daniel Straub}
\affiliation[label3]{organization={Engineering Risk Analysis Group \& Munich Data Science Institute, Technical University of Munich},
            addressline={Theresienstrasse 90}, 
            city={Munich},
            postcode={80333}, 
            country={Germany}}

\title{Near-real-time ship grounding damage assessment using Bayesian networks} 

\begin{abstract}
In a post-grounding event, the rapid assessment of hull girder residual strength is crucial for making informed decisions, such as determining whether the vessel can safely reach the closest yard. One of the primary challenges in this assessment is the uncertainty in the estimation of the extent of structural damage. Although classification societies have developed rapid response damage assessment tools, primarily relying on 2D Smith-based models, these tools are based on deterministic methods and conservative estimates of damage extent. To enhance this assessment, we propose a probabilistic framework for rapid grounding damage assessment of ship structures using Bayesian networks (BNs). The proposed BN model integrates multiple information sources, including underwater inspection results, hydrostatic and bathymetric data, crashworthiness models, and hydraulic models for flooding and oil spill monitoring. By systematically incorporating these parameters and their associated uncertainties within a causal framework, the BN allows for dynamic updates as new evidence emerges during an incident. Two case studies demonstrate the effectiveness of this methodology, highlighting its potential as a practical decision support tool to improve operational safety during grounding events. The results indicate that combining models with on-site observations can even replace costly underwater inspections.
\end{abstract}

\begin{keyword}
Ship grounding \sep Rapid damage assessment \sep Bayesian networks \sep Residual strength \sep Decision support

\end{keyword}

\end{frontmatter}

\section{Introduction} 
Ship grounding accounts for about one third of commercial ship accidents \citep{Sam2009,Eliopoulou2016}. Their severity depends on many uncertain factors, including seafloor morphology, environmental conditions and operational aspects. In a favorable scenario, grounding may occur on a sandy seabed at low speed, resulting in minor damage. In a worst-case scenario, grounding can cause hull breach, which can lead to water ingress and oil spill. In this type of grounding, the ship is usually moving forward by its speed (powered grounding) and the morphology of the seabed is rocky. Such situations jeopardize the structural integrity and stability of the vessel and can ultimately result in its total loss and human casualties \citep{ISSC2022ALS}. 

Damage (or hull breach) extent identification is crucial in the assessment of ship survivability, both in terms of residual strength \citep{Fang2005,Parunov_2017,Βuvzanvcic2020} and stability \citep{Ruponen2017,Zhang2021,Hirdaris2023}. Commonly, the actual damage extent is idealized by a rectangular-shaped (2D) or a box-shaped (3D) geometry. The uncertainty in the estimation of the extent of bottom structural damage following a grounding event asks for the use of probabilistic methods for its description. In the past, statistical data from historical accidents were used to establish global-based distribution models for the representation of the extent and location of damage \citep{IMO2003,Zarafonitis2020}. Although the adoption of these global-based distribution models is sufficient for design considerations \citep{IACS2024}, it is less relevant for the assessment of hull-girder residual strength in real conditions.

In a real grounding scenario, a quick and accurate estimation of the damage extent is needed for the evaluation of hull-girder residual strength. In this direction, various methods have been proposed. \citet{Nguyen2011} presented a simple and quick procedure to estimate the damage to the hull of a ship in a powered grounding event based on numerical analysis. \citet{Pineau2022} developed a numerical tool for ship grounding damage assessment that couples a super-element solver with the 6-DoF external dynamics. Both studies, however, are intrinsically deterministic and they are exclusively based on energy considerations. A probabilistic method to estimate damage extent following the grounding of a passenger ship was proposed by \citet{Hirdaris2023}. Ship operating conditions before grounding were modeled using distribution models based on statistics of past events and Monte Carlo simulation was used to generate a set of plausible damage extents. This method, however, does not allow for the updating on input uncertain parameters using evidence from the incident site.

Classification societies have developed dedicated tools for rapid response damage assessment, such as the RRDA tool of ABS \citep{ABS2012} or the ERS software of DNV \citep{ERS}. In these tools, the estimation of the damage extent is typically based on the information provided by the master, the divers' inspection, or both. However, due to incomplete and limited data, conservative estimates are generally adopted; for example, damage estimates are increased to approximately two to five times the reported value \citep{ABS2012}. These tools do not explicitly consider the uncertainty in the basic parameters of the problem and the subsequent damage estimation through probabilistic terms.

To fill the above gaps, we propose the development of a probabilistic graphical model implemented in the form of a Bayesian network. Bayesian networks (BNs) are powerful tools whose purpose is to give probabilistic estimates for events that are not directly observable or partly observable (hypothesis variables) by providing some information channels (information variables) that may reveal something about the hypothesis variables \citep{Jensen2007}. BNs have been used for various applications in the maritime industry [see e.g., \citet{Hansen2000,Animah2024}]. With respect to ship grounding, BNs have been mostly applied for the evaluation of the occurrence probability of a grounding event and its associated risk assessment \citep{Mazaheri_2016,Jiang_2021}. A work towards hull-girder residual strength assessment was presented by \citet{Li_2019}, who developed a structural reliability analysis model based on BNs for quantifying the collapse risk of damaged hulls after grounding. None of the aforementioned studies, however, addresses damage assessment after grounding. 

In this paper, we present a BN model for predicting the extent of damage after hard grounding events on oil tankers. The proposed BN makes use of available physics-based models (i.e., crashworthiness and hydraulic flow models) and on-site information (i.e., inflow or outflow monitoring, draft measurements, stability data, bathymetric data and underwater inspection outcomes) in a formal and intuitive way.  We demonstrate the robustness of the proposed BN model, first, by providing a validation for parts of the model using a real grounding case of a single-hull tanker, and second, by examining a hypothetical grounding scenario on a double-hull VLCC tanker. We conclude this paper by discussing the implications and potential application of the model in real-life incidents, as well as the limitations and future extensions of this framework. To the authors' knowledge, this is the first study that provides a probabilistic framework for rapid assessment of damage following grounding in real operational conditions.

\section{Problem statement}     \label{Problem}

\subsection{Background}     \label{Background}
The inaccessibility of the submerged bottom structure during a grounding event poses significant challenges to the identification of the damage extent. Direct evidence is only available through underwater inspection from divers; however, such inspections are not always feasible (e.g., due to bad weather conditions or high currents) and may involve significant uncertainties (e.g., due to poor visibility). In these cases, the damage assessment is typically based on indirect information collected on board, such as draft and water depth measurements, flooding or oil outflow rate monitoring, and oil-water mixture detection, among others. Additional insights into the damage extent can be acquired from crashworthiness and hydraulic flow models. 

The structural crashworthiness of ships during grounding events has been analyzed using finite element analysis (FEA) and semi-analytical (or empirical) methods. Although FEA allows for a detailed investigation of the impact process \citep{Kitamura2002,Paik2007a,Paik2007b,Abubakar2013,Liu2021,Alsos2007,Liu2018,Prabowo2019}, its use is dictated by significant computational time and modeling effort that is prohibitive for real-time applications. On the other hand, semi-analytical formulas are practical and can provide a first-cut estimate of the damage extent. These formulas are primarily relying on energy considerations; they connect the kinetic energy of the ship (speed and displacement) and its structural resistance (double bottom structural layout and building material) to the extent of grounding damage. Several well-known formulas have been developed in the past [e.g., \citet{Pedersen2000,Zhang2002,Hong2012}]. In more recent years, new empirical formulas have been proposed which have been calibrated using advanced FEA simulations [e.g., \citet{cerup2009, Heinvee2015}]. Due to their simplicity, advanced empirical formulas will be used in this paper.

Hydraulic flow models are typically used for direct consequence analysis; for a given damage extent, the flooding rate or the total oil spill volume and duration is calculated \citep{Sergejeva2017,Tabri2018,Kollo2017}. When combined with on-board measurements, these models can offer valuable insights into the damage extent by means of an inverse analysis. \citet{Ruponen2017}, for example, has demonstrated how hydraulic flooding models in association with flood level sensors located inside compartments can be used to estimate the size of the damage area. This form of inverse analysis will be utilized in this paper.

\subsection{Proposed method}    \label{Proposed method}
\begin{figure}
    \centering
    \includegraphics[width=0.75\linewidth]{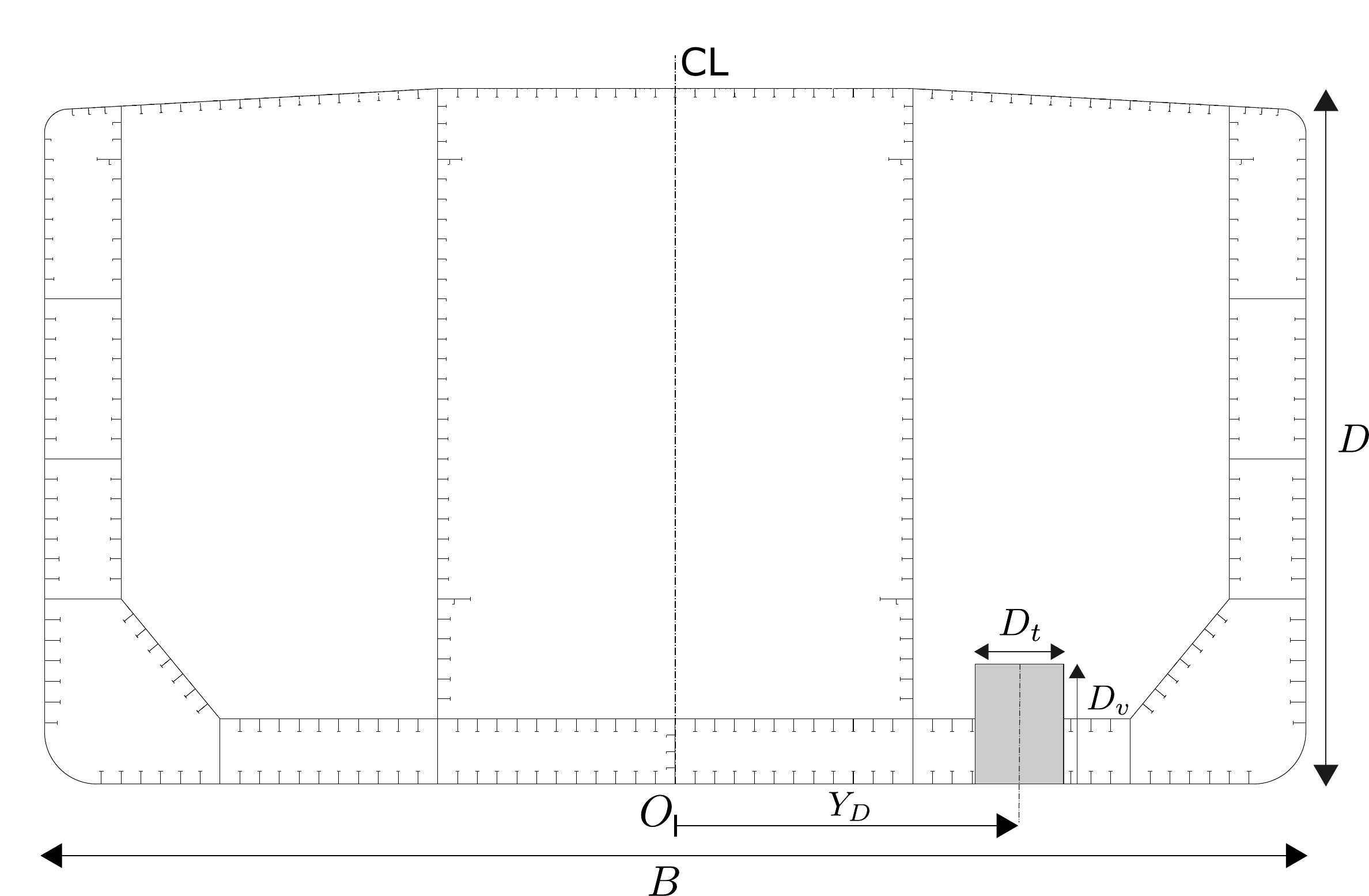}  
    \caption{Bottom damage schematics in the cross-section.}
    \label{fig:Grounding schematics}
\end{figure}

The purpose of this paper is to provide a method for estimating probabilistically the 2D damage extent and location. We describe the damage by the following three parameters, shown in Figure \ref{fig:Grounding schematics}: the transverse location of the center of the damage $Y_D$, the extent of the transverse damage $D_t$, and the extent of the vertical damage (penetration) $D_v$. These parameters are modelled as random variables. In this work, we consider the longitudinal variables (longitudinal extent and location) of the damage to be known with limited uncertainty. The elimination of longitudinal variables is to be seen in accordance with the adoption of the standard 2D Smith's model adopted by \cite{IACS2024}, which is typically used for hull-girder residual strength analysis in real conditions \citep{ABS2012}.

In the first step of the proposed methodology, we identify various sources of information that provide insights on the damage extent, including (i) crashworthiness empirical formulas, (ii) hydraulic flow models (flooding and oil spill models) and (ii) in-situ information (hydrostatics and stability data, bathymetric data and underwater inspection). In a second step, this information is supplemented with experts' judgment to develop a suitable BN model for estimating damage extent in near-real-time conditions. The BN integrates and formalizes all of this information and its associated uncertainty in a causal and compact manner and allows for the updating of parameters as new evidence comes into light at the time of the incident. During an incident, the BN will enable rapid (near-real-time) probabilistic assessment of the extent of damage by providing the probability distribution of $Y_D$, $D_t$ and $D_v$. This information can then serve as input to a subsequent probabilistic residual strength assessment.

\subsection{Assumptions}        \label{Limitations}
In a grounding scenario, the seabed topologies are generally classified into three categories: rock, reef, and shoal \citep{Alsos2007}. Here, we examine the consequences of hard grounding (rock scenario), which causes extensive crushing and tearing damage, when the ship is powered by its own engine; the ship is not drifted by currents or waves to a rock. 

An idealized rectangular-shaped damage area is assumed (see Figure \ref{fig:Grounding schematics}). This implies that the damaged widths of the outer and inner shells are the same. Although this may not be the real case, it is common to assume that all damaged structures, regardless of the grounding damage modes (e.g., tearing and denting), are completely ineffective in residual strength calculations \citep{Li2022}. The simplified rectangular-shaped area can produce sufficiently accurate results and provides a rational conservatism on residual strength assessment \citep{Jiang2014}.

Finally, we assume that the critical damage is located within the parallel body of the ship’s hull, where longitudinal bending moments generally reach their maximum.

\section{Bayesian networks}         \label{BNs theory}
\subsection{Modelling}
The Bayesian network (BN) is a directed acyclic probabilistic graphical model formed by nodes and direct links among the nodes \citep{Pearl1995,Jensen2007}. The nodes in the graph represent random variables and the links encode the dependence structure among these random variables. Ideally, the link between two nodes is based on a causal relation, but this is not necessary. Each random variable $X_i$ is characterized by its conditional probability distribution $p(x_i|pa(x_i))$ given its parents $pa(x_i)$, which is the set of nodes with links pointing toward $X_i$. In the case of discrete random variables, this conditional distribution takes the form of conditional probability tables (CPTs). In the case of continuous random variables, the conditional distribution is characterized by a conditional probability density function (PDF). A BN in which all its random variables are discrete is termed \textit{discrete} BN. If there is a mix of discrete and continuous random variables, the BN is called \textit{hybrid} BN. 

Any BN specifies the joint probability distribution $p(\mathbf{x})$ of a set of random variables as the product of the conditional distributions:

\begin{equation} \label{BN pdf joint}
    p(\mathbf{x}) = \prod_{i=1}^{n}p[x_i|pa(x_i)]
\end{equation}

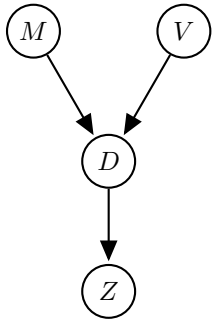
\begin{figure}[h]
  \centering
    \begin{tikzpicture}[shorten >= 1pt, auto, thick]
        
    \node[latent] (Damage) {$D$};
    \node[latent, above=1.0 of Damage, xshift=-1.0cm] (Mass) {$M$};
    \node[latent, above=1.0 of Damage, xshift=+1.0cm] (Speed) {$V$};   
    \node[latent, below=1.0 of Damage] (Survey) {$Z$};

    \edge {Mass} {Damage}
    \edge {Speed} {Damage}
    \edge {Damage} {Survey}

    \end{tikzpicture}
\caption{A simple BN where $M$ represents the mass of the ship, $V$ is the impact speed, $D$ is the damage and $Z$ is the inspection outcome.}
\label{fig:BN Example}   
\end{figure}

As an example, consider the BN of Figure \ref{fig:BN Example}. The BN represents a simplified version of the crashworthiness model (defined with more detail in Section \ref{BN Energy}). A ship characterized by its mass $M$ and speed $V$ strikes a rock. A damage $D$ is caused on its bottom structure. An additional node $Z$ is included representing the outcome of an inspection. The nodes in this example represent discrete random variables whose joint PMF is:

\begin{equation} \label{BN pdf example}
    p(m,v,d,z) = p(z|d) p(d|m,v) p(m) p(v)   
\end{equation}

The links in the BN provide information on the dependence between random variables in the model. For example, $M$ and $V$ are assumed to be independent a-priori, and hence no direct link between them is present. The link from $D$ to $Z$ reflects that the inspection result depends on the damage state. It provides no direct information on $M$ and $V$. However, it does so indirectly, because the information obtained on $D$ also updates the probability distribution of $M$ and $V$, as long as $D$ is not known with certainty. In this way, by observing one random variable, potentially all others are updated.

\subsection{Inference}
Using BNs it is possible to obtain the posterior distribution of a set of random variables given a set of observations (or evidence, $\mathbf{e}$). This task is called \textit{inference}. For instance, if an inspection result is included in the previously presented example, i.e., $Z=e$, then the (joint) probability distribution of the random variables $M$, $V$ and $D$ conditional on the observed value of $Z$ is calculated as:

\begin{equation} \label{BN inference example}
    p(m,v,d|Z=e) = \frac{p(m,v,d,e)}{p(e)}   
\end{equation}

A variety of algorithms exist for inference, which can be broadly classified into two main categories: \textit{exact} and \textit{sampling-based} algorithms. For efficient computation, all BN inference algorithms make use of the graphical structure by performing computations locally, exploiting the conditional independence assumptions encoded in the graph.

In this work, we employ an exact algorithm called \textit{clustering algorithm} \citep{Lauritzen1988}. Since the BN considered in this study consists of both discrete and continuous random variables, applying exact inference first requires the discretization of the continuous variables. More specifically, the original continuous domain of each random variable is partitioned into discrete intervals and the probability of each interval is computed from the conditional or the marginal PDF of the random variable. Even though the inference algorithms are exact for a given discretization, the discretization itself introduces an error. The number and location of the discrete intervals have an impact on the computation time and accuracy of the approximation.

\section{Bayesian network model for damage assessment}      \label{BN development}
In this section, we develop the BN model for 2D damage assessment. In Section \ref{Damage characterization}, we perform the discretization of the nodes that characterize the 2D extent of damage (nodes of interest). In the following sections, we develop the basic modules (information variables) of the network and their dependence structure step by step. The full BN model is presented in Section \ref{Full BN}.

\subsection{Damage opening description}    \label{Damage characterization}
In the BN model, the 2D damage extent and location are represented by three continuous random variables, $Y_D$, $D_t$ and $D_v$ (see Section \ref{Proposed method}). The plausible ranges of these variables are presented in Table \ref{table:Damage description}. The total number of states will vary depending on the values of ship breadth $B$ and depth $D$. For transverse damage extent and location, an interval size equal to 1 m is considered sufficient for the discretization. A more coarse discretization scheme can be used for larger damage widths.

\begin{figure}
    \centering
    \includegraphics[width=1.0\linewidth]{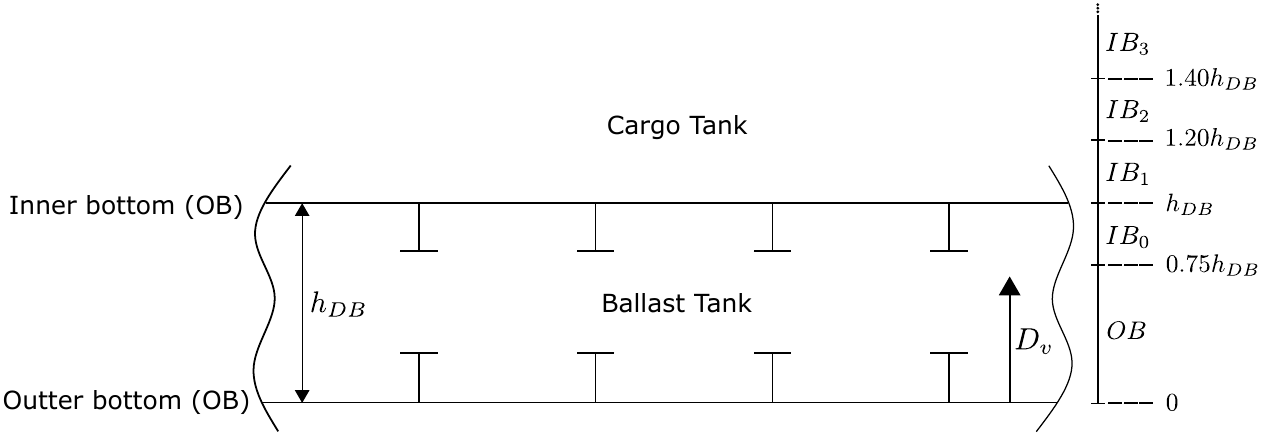}
    \caption{Assigned states for vertical damage extent $D_v$.}
    \label{fig:Penetration extent}
\end{figure}

The vertical damage extent is discretized to a number of states as a function of double bottom height $h_{DB}$ (see Figure \ref{fig:Penetration extent}). In the first state, OB, the penetration depth is assumed to reach about 3/4 of double bottom height. Within this height, the bottom outer shell, the attached girders and longitudinal stiffeners are considered ineffective. In the second state, IB0, the penetration depth is assumed to reach the double bottom height, i.e., the attached girders and longitudinal stiffeners of the inner bottom are considered ineffective. The inner hull breach does not necessarily occur when the vertical damage extent reaches the double bottom height; \citet{Heinvee2015} showed that the critical penetration for inner hull breach is about $1.4h_{DB}$. Therefore, reaching the states IB1 or IB2 may not cause inner bottom hull breach -- although severe damage in the form of large plastic deformation and denting is expected to the inner shell and the attached stiffeners and girders. As the penetration extent increases, inner bottom breach becomes more likely. To quantify this phenomenon, the discrete node ``IHB" (inner hull breach) is introduced conditional on $D_v$ and the CPT shown in Table \ref{tab:IBB} is formulated. The probabilities assigned in Table \ref{tab:IBB} are estimated based on the relative frequencies presented by \citet{Heinvee2015}.

\begin{table}[]
\centering
\caption{Plausible ranges for the three parameters describing the 2D damage extent and location. They are all represented by continuous random variables in the BN model.}
\label{table:Damage description}
\begin{tabular}{lccc}
\hline
Parameter                   & Symbol    & Plausible range   & Units  \\ \hline
Vertical extent             & $D_v$     & {[}0, 0.3D{]}     & m      \\
Transverse extent           & $D_t$     & {[}0, B{]}        & m       \\
Transverse center location  & $Y_D$     & {[}-B/2, B/2{]}   & m       \\ \hline
\end{tabular}
\end{table}

\begin{table}[]
\centering
\caption{CPT showing the probability of inner hull breach IHB given the state of vertical damage extent $D_v$. The probabilities are estimated based on the results presented by \citet{Heinvee2015}.}
\label{tab:IBB}
\begin{tabular}{lccccccc}
\hline
\multicolumn{2}{l}{Vertical damage extent, $D_v$} & OB & IB0 & IB1 & IB2 & IB3 & IB4... \\ \hline
\multirow{2}{*}{Inner hull breach}    & yes   & 0  & 0   & 0.7 & 0.9 & 0.95 & 1     \\
                                      & no    & 1  & 1   & 0.3 & 0.1 & 0.05 & 0    \\ \hline
\end{tabular}
\end{table}

\subsection{Basic modules}      \label{Basic modules}
Based on a literature survey and engineering judgment, we identified key variables that can provide insights into the extent and location of damage. These variables are grouped into four basic modules, which are presented in the following sections. These modules are the: (i) Crashworthiness module, (ii) Hydraulic module, (iii) Hydrostatics and bathymetric module, and (iv) Inspection module.

\subsection{Crashworthiness module}     \label{Crashworthiness model}
The crashworthiness module is based on empirical models that connect the kinetic energy of the ship before grounding and its (double) bottom structural resistance to the extent of grounding damage. The kinetic energy $E$ of the ship before grounding is:

\begin{equation} \label{E}
    E = \frac{1}{2}(M + M_a)V^2
\end{equation}
where $M$ denotes the mass (total displacement or weight) of the ship, $M_a$ is the added mass, usually taken equal to 5\% of $M$ \citep{Pedersen2000,Zhang2002} and $V$ is the impact speed of the vessel. 

A common assumption is that the kinetic energy of the ship before grounding is totally dissipated by the destruction of the ship's bottom structure \citep{Pedersen2000,Zhang2002}. Thus, the horizontal grounding force $F_H$ can be estimated as:

\begin{equation} \label{Fh1}
    F_H = \frac{0.5(M + M_a)V^2}{L_D} \times \epsilon_{f_h}
\end{equation}
where $L_D$ is the damage (stopping) length and $\epsilon_{f_h}$ is a multiplicative error term that follows the lognormal distribution with a median equal to one and coefficient of variation $\delta_{\epsilon_{f_h}}=0.10$ \citep{cerup2009}. This error reflects the uncertainty in the grounding force prediction. 

The horizontal grounding force depends on the structural arrangement of the double bottom, the material, and the width of the rock (or equivalently, the damage width). As mentioned in Section \ref{Background}, several formulas have been developed for the estimation of total horizontal grounding force in rocky morphology, where plate tearing is the dominant failure mode. Here, we use the empirical formula developed by \citet{cerup2009}, which has been calibrated with a series of FEA and validated against real grounding scenarios. According to this formula, the horizontal grounding force $F_H$ can be estimated as:

\begin{equation} \label{Fh2}
    F_H=0.77\sigma_0\epsilon_{f}^{0.71}t_{eq}^{1.17}D_t^{0.83}
\end{equation}
where $\sigma_0$ is the flow stress of the outer bottom, given as the average between the yield stress and the ultimate stress of steel, $\epsilon_f$ is the fracture strain and $t_{eq}$ is the equivalent plate thickness. To predict the damage width conditional on the horizontal grounding force, Eq. \ref{Fh2} is rearranged to solve for $D_t$:

\begin{equation} \label{Dt}
    D_t = \left[ \frac{F_H}{0.77 \left( \underbrace{\sigma_0\epsilon_{f}^{0.71}t_{eq}^{1.17}}_{\rm{Outer~bottom}} + \underbrace{\sigma_0\epsilon_{f}^{0.71}t_{eq}^{1.17}}_{\rm{Inner~bottom}} \right) }\right] ^{\sfrac{1}{0.83}}
\end{equation}
Eq. \ref{Dt} takes into account both the outer and inner hull resistance. If only the outer bottom is damaged, the inner bottom terms are neglected.

\subsubsection{Impact speed}
The impact speed of the vessel on the rock $V$ is modeled as a continuous random variable. The prior density function can be specified based on historical data from past accidents. Based on the findings of \citet{cerup2009} and \citet{Youssef2018}, a Beta distribution with parameters $\alpha=5$ and $\beta=2$ is adopted here (see Figure \ref{fig:V prior}). Evidence can be provided for the reported speed $V_r$, which is used to update our prior belief. It is assumed that the following relationship holds:

\begin{equation} \label{V}
    V_r = V + \epsilon_v
\end{equation}
where $V$ is the true impact speed and $\epsilon_v$ denotes the error in the reported value, which is modelled by a normal distribution with zero mean and standard deviation $\sigma_{\epsilon_v}=0.24$ kn \citep{Dalheim2021}.

\subsubsection{Displacement}
The displacement of the ship before the accident is modeled as a uniform distribution with lower and upper limits defined among the various seagoing loading conditions of the ship. Evidence can be provided for the reported displacement $M_r$, which can be used to update our initial belief. It is assumed that the following relationship holds:

\begin{equation} \label{M}
    M_r = M \times \epsilon_m
\end{equation}
where $M$ is the true displacement and $\epsilon_m$ is the error in the reported value, which is modelled by a lognormal distribution with median one and coefficient of variation $\delta_{\epsilon_m}=2.5\%$ \citep{ClassNK2014}.

\subsubsection{Damage length}
\begin{figure}[h]
    \centering
    \begin{subfigure}{0.45\textwidth}
    \centering
    \includegraphics[width=\textwidth]{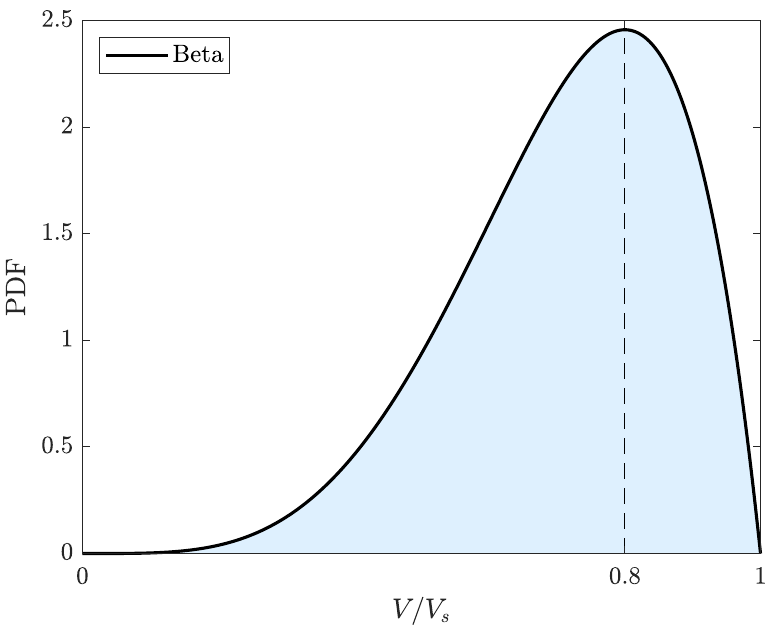}
    \caption{}
    \label{fig:V prior}
    \end{subfigure}
    \hfill
    \begin{subfigure}{0.45\textwidth}
    \centering
    \includegraphics[width=\textwidth]{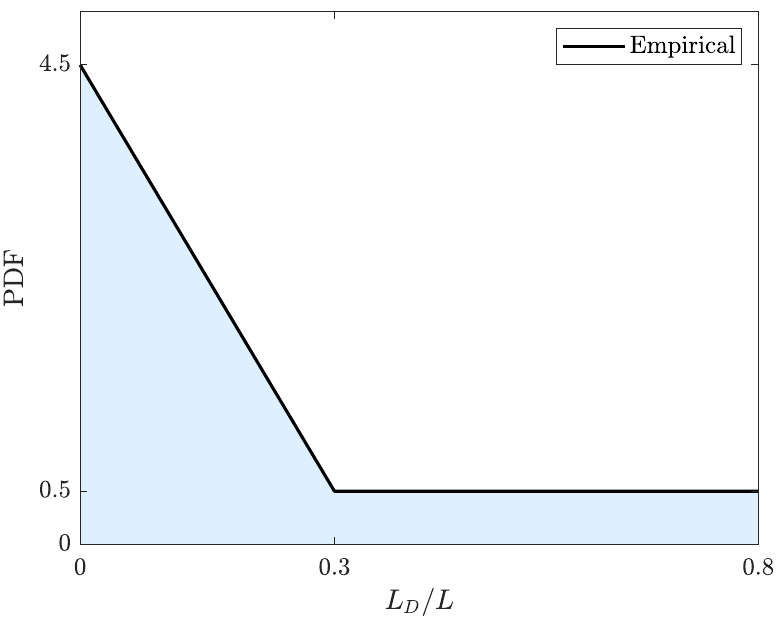}
    \caption{}
    \label{fig:Ld prior}
    \end{subfigure}

    \caption{Prior distribution models for normalized impact speed and damage length.}
    \label{fig:image1}
\end{figure}

Information on the damage length extent can be acquired from past accidents. The distribution depicted in Figure \ref{fig:Ld prior} is used as a prior distribution in our BN model \citep{IMO2003}. Evidence on the reported damage length $L_{D,r}$ on-site can be used to update our prior belief. The following relationship is assumed:

\begin{equation} \label{Ld}
    L_{D,r} = L_D + \epsilon_l
\end{equation}
where $L_D$ is the true damage length and $\epsilon_l$ denotes the error in the reported value, which is modelled by a normal distribution with zero mean and standard deviation $\sigma_{\epsilon_l}$. The deviation of the reported value from the actual value will depend on the information provided after the accident (e.g., reported damaged tanks). For the current BN model, a value equal to $\sigma_{\epsilon_l}=5$ m is considered.

\subsubsection{BN model}    \label{BN Energy}
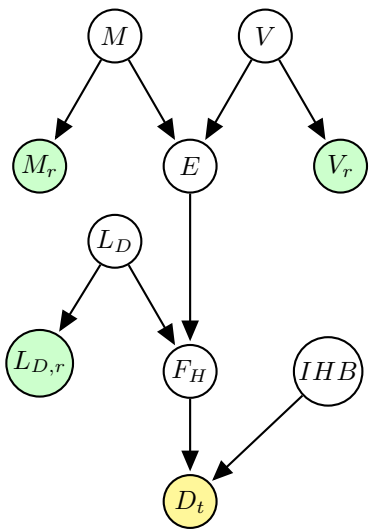
\begin{figure}[t]
  \centering
    \begin{tikzpicture}[shorten >= 1pt, auto, thick]
        
    \node[latent, xshift=-1cm] (Mass) {$M$};
    \node[latent, fill=green!20, below=1.0 of Mass, xshift=-1cm] (Mass rep) {$M_r$};
    \node[latent, xshift=+1cm] (Speed) {$V$};   
    \node[latent, fill=green!20, below=1.0 of Speed, xshift=+1cm] (Speed rep) {$V_r$};
    \node[latent, below=1.0 of Mass, xshift=+1cm] (Energy) {$E$};
    \node[latent, below=2.0 of Energy] (Force hor) {$F_H$};
    \node[latent, below=2.0 of Mass] (Damaged length) {$L_D$};
    \node[latent, fill=green!20, below=0.8 of Damaged length, xshift=-1cm] (Damaged length rep) {$L_{D,r}$};
    \node[latent, fill=yellow!50, below=1.0 of Force hor] (Damage t) {$D_t$};
    \node[latent, right=1.0 of Force hor] (IHB) {$IHB$};
    
    \edge {Mass} {Mass rep}
    \edge {Mass} {Energy}
    \edge {Speed} {Speed rep}
    \edge {Speed} {Energy}
    \edge {Energy} {Force hor}
    \edge {Damaged length} {Force hor}
    \edge {Damaged length} {Damaged length rep}
    \edge {Force hor} {Damage t}
    \edge {IHB} {Damage t}

    \end{tikzpicture}
\caption{BN for the prediction of damage width extent $D_t$ based on energy considerations. Green shaded nodes indicate inputs and yellow indicates the node of interest.}
\label{fig:BN Energy}   
\end{figure}

The full BN that connects the initial kinetic energy of the ship to the damage width extent $D_t$ is illustrated in Figure \ref{fig:BN Energy}. A validation of that model against a real accident case is provided in Section \ref{Case study 1}.

\subsection{Hydraulic module}       \label{Hydraulic models}
The hydraulic module consists of a hydraulic flow model connecting the monitored flooding or oil spill rate with the extent of damage.

\subsubsection{Water ingress/oil outflow detection}
The detection of oil outflow or water ingress inside a cargo tank offers a key information: whether the inner bottom shell has been breached or not. For instance, when an oil tanker is loaded, water ingress detection in the ballast tank implies outer bottom breach, while oil outflow implies inner bottom breach. The CPTs of Tables \ref{table:CPT water ingress} and \ref{table:CPT oil spill} quantify this effect. Note that we assume here that our knowledge for detecting oil outflow or water ingress is perfect; water ingress can be detected via tank soundings or water level sensors, while oil outflow can be detected through tank ullages or visual observation at sea (oil spill).

\begin{table}[]
\centering
\caption{CPT of water ingress (WI) detection in ballast or cargo tank area conditional on the inner hull breach and the loading condition, $p(wi|ihb,lc)$.}
\label{table:CPT water ingress}
\begin{tabular}{lccccc}
\hline
\multicolumn{2}{l}{Loading condition, LC}    & \multicolumn{2}{c}{Loaded} & \multicolumn{2}{c}{Ballast} \\ \hline
\multicolumn{2}{l}{Inner hull breach, IHB} & yes          & no          & yes           & no          \\ \hline
\multicolumn{1}{c}{\multirow{2}{*}{Water ingress}}                & Ballast tank               & 1            & 1           & 0             & 1           \\
                         & Cargo tank                & 0            & 0           & 1             & 0           \\ \hline
\end{tabular}
\end{table}

\begin{table}[]
\centering
\caption{CPT of oil outflow detection conditional on the inner hull breach and the loading condition, $p(os|ihb,lc)$.}
\label{table:CPT oil spill}
\begin{tabular}{lccccc}
\hline
\multicolumn{2}{l}{Loading condition, LC}    & \multicolumn{2}{c}{Loaded} & \multicolumn{2}{c}{Ballast} \\ \hline
\multicolumn{2}{l}{Inner hull breach, IHB} & yes          & no          & yes           & no          \\ \hline
\multicolumn{1}{c}{\multirow{2}{*}{Oil outflow}}                & yes               & 1            & 0           & 0             & 0           \\
                         & no                & 0            & 1           & 1             & 1           \\ \hline
\end{tabular}
\end{table}

\subsubsection{Inflow/outflow rate formulas}
The inflow rate of sea water or the outflow rate of oil depends on the size of the damage opening. Therefore, monitoring these rates provides information on the damage area. Hydraulic flow models based on Bernoulli’s principle establish a direct relationship between flow rate and the size of the damage opening.

\begin{figure}[h]
    \centering
    \begin{subfigure}{0.45\textwidth}
    \centering
    \includegraphics[width=\textwidth]{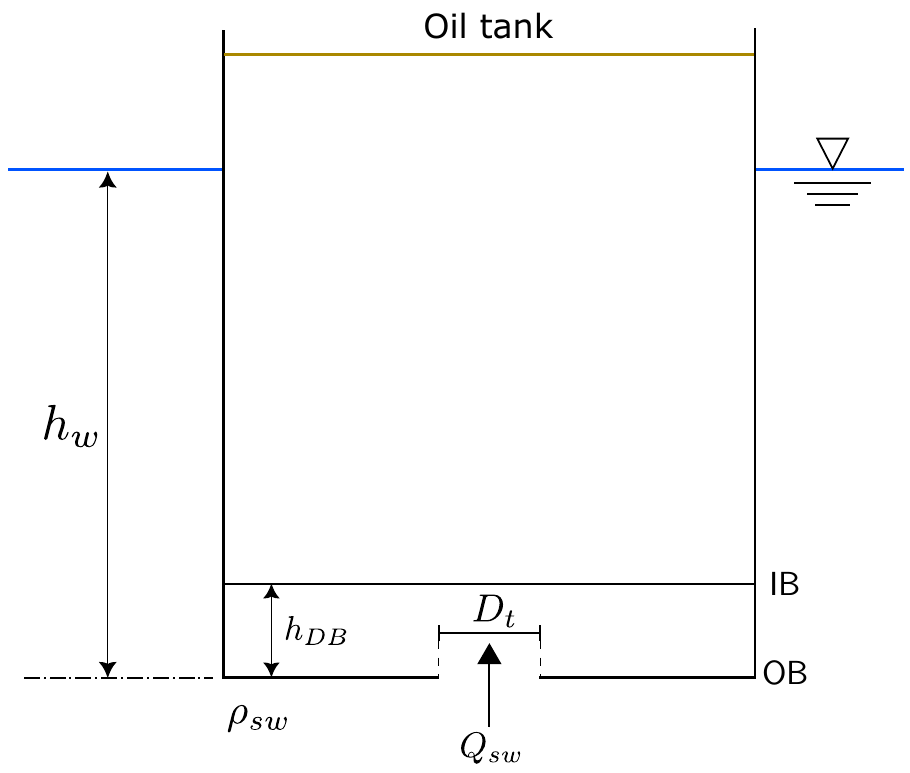}
    \caption{Water ingress.}
    \label{fig:Water ingress}
    \end{subfigure}
    \hfill
    \begin{subfigure}{0.45\textwidth}
    \centering
    \includegraphics[width=\textwidth]{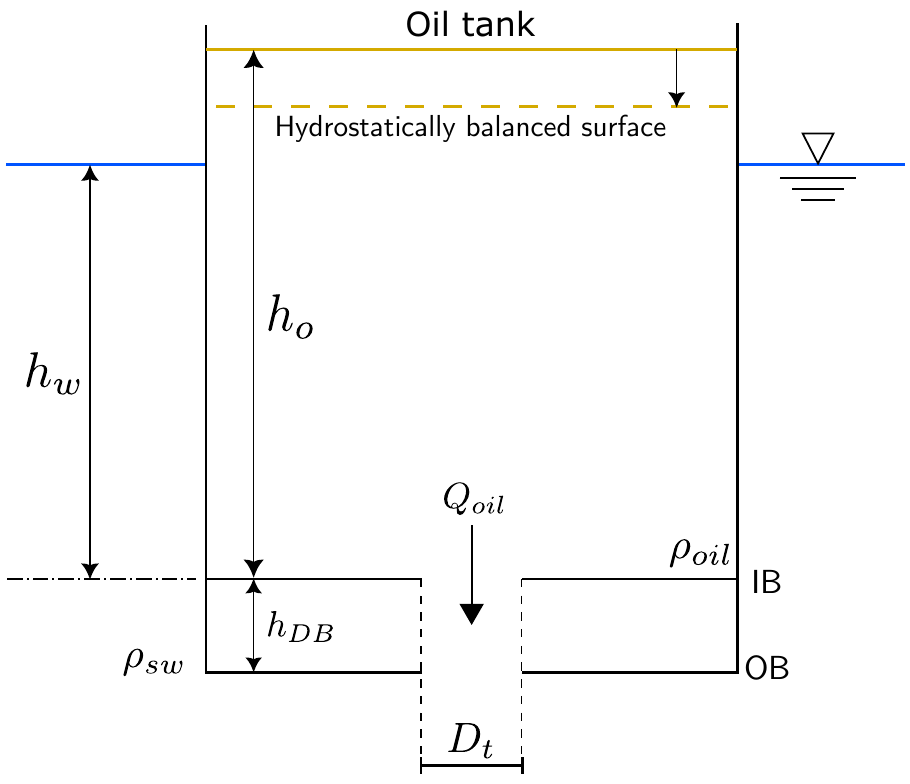}
    \caption{Oil spill.}
    \label{fig:Oil spill}
    \end{subfigure}

    \caption{Sketches for the unidirectional water ingress flow (left) and oil outflow (right) through the bottom hull and inner bottom hull opening and notations.}
    \label{fig:image2}
\end{figure}

The flooding rate $Q_{sw}$ through a damage opening can be obtained using the following formula \citep{Ruponen2017}:

\begin{equation} \label{Qsw}
    Q_{sw} = C_d  A \sqrt{2gh_w}
\end{equation}
where $C_d$ is a discharge coefficient, $g$ is the acceleration of gravity, $A$ is the area of the damage opening and $h_w$ is the effective pressure head (i.e., the vertical distance from the damage opening to the sea surface level) evaluated based on the monitored draft, trim and heel (see Figure \ref{fig:Water ingress}).

The discharge coefficient is usually set to 0.60 \citep{Ruponen2017} or 0.65 \citep{Tavakoli2008}. Here, we assign these values to the 15\% and 85\% quantiles and we fit a normal distribution. The associated mean and standard deviation for $C_d$ are thus 0.625 and 0.02.

Assuming a box-shaped damage geometry (i.e., constant damage width throughout the damage length), the damage area is $A=l_D D_t$, where $l_D$ is the damage length of the tank. The uncertainty in $l_D$ can be eliminated if the damage has propagated to the entire tank's length (in this case, $l_D$ equals the length of the tank). If the damage length is uncertain, an additive zero-mean Gaussian error can be assumed. 

The oil outflow rate $Q_{oil}$ through the inner hull opening can be calculated, similarly to Eq. \ref{Qsw}, using the following relationship \citep{Tavakoli2008,Sergejeva2013}:

\begin{equation} \label{Qoil}
    Q_{oil} = C_d  A \sqrt{2g(h_o - \frac{\rho_w}{\rho_o}h_w)}
\end{equation}
where $h_w$ is the vertical height from the sea surface to the inner bottom opening, $h_o$ is the initial level of oil inside the tank, $\rho_o$ is the oil density and $\rho_w$ is the sea water density (see Figure \ref{fig:Oil spill}).

For both Eq. \ref{Qsw} and \ref{Qoil}, evidence on $Q_{sw}$ or $Q_{oil}$ allows for inference about $D_t$ using Bayesian updating. The appropriate formula is used based on the ship's loading condition state (i.e., loaded or ballast) and the inner hull breach (i.e., breach or no breach).

\subsubsection{Volumetric flow measurement}
The volumetric inflow or outflow rate $Q$ can be evaluated using the central difference method as:

\begin{equation} \label{DV_Dt}
    Q(t) = \frac{dV(t)}{dt} = \frac{V(h(t) + \frac{\Delta t}{2}\frac{dh(t)}{dt}) -  V(h(t) - \frac{\Delta t}{2}\frac{dh(t)}{dt})}{\Delta t}
\end{equation}
where $V(h)$ is the volume of ingress (oil lost) for a given water (oil) level $h$ (considering the 3D geometry of the tank), $dh/dt$ is the measured level rate in the tank and $\Delta t$ is the measured time step, typically about 2.0 s (0.50 Hz) to 5.0 s (0.20 Hz). A measurement data period of about 40 s to 60 s is generally recommended \citep{Ruponen2017}.

\begin{figure}[h]
  \centering
    \begin{tikzpicture}[shorten >= 1pt, auto, thick]
        
    \node[latent] (Inner hull breach) {$IHB$}; 
    \node[latent, fill=green!20, below=1.0 of Inner hull breach, xshift=+0.5cm] (Oil spill) {$OS$};
    \node[latent, fill=green!20, right=1.0 of Oil spill] (Water ingress) {$WI$};  
    \node[latent, fill=green!20, above=1.0 of Water ingress] (Loading condition) {$LC$};  
    \node[latent, below=3.0 of Inner hull breach] (Flow rate) {$Q$}; 
    \node[latent, fill=green!20, below=1.0 of Flow rate] (Flow rate m) {$Q_m$};
    \node[latent, fill=green!20, left=1.0 of Flow rate m] (Q error) {$Q_{\epsilon}$};
    \node[latent, fill=yellow!50, left=1.0 of Flow rate] (Damage t) {$D_t$};
       
    \edge {Inner hull breach} {Oil spill}  
    \edge {Inner hull breach} {Water ingress}  
    \edge {Loading condition} {Oil spill}
    \edge {Loading condition} {Water ingress}    
    \edge {Inner hull breach} {Flow rate}
    \edge {Loading condition} {Flow rate}
    \edge {Damage t} {Flow rate}
    \edge {Flow rate} {Flow rate m}
    \edge {Q error} {Flow rate m} 

    \end{tikzpicture}
\caption{BN for vertical and transverse damage extent prediction based on water ingress and oil spill detection and monitoring. The quality of flow rate measurement is represented by the discrete node $Q_\epsilon$.}
\label{fig:Water ingress and oil spill}   
\end{figure}
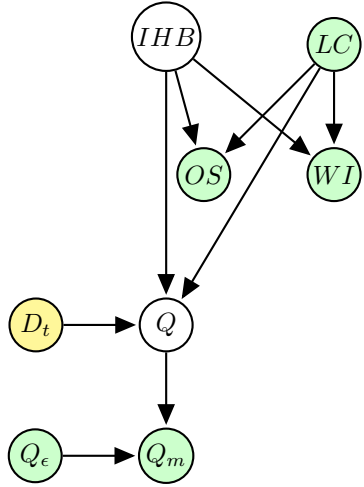

The analysis of the damage extent should be performed soon after flooding has been detected because the effective pressure head and the flow rate decrease as the flooding progresses \citep{Ruponen2017}. Several sources of uncertainty may affect the measurements, including sloshing of liquids inside tanks and measurement tool noise. Measurements can be obtained manually; the increase in water level can be measured by taking tank soundings, while the decrease in oil level can be measured by taking tank ullages. More accurate predictions are possible if level sensors are installed inside the tanks. In both cases, we assume a multiplicative error term in the measured flow rate of the form:

\begin{equation} \label{Qm}
    Q_m = Q \times \epsilon_q
\end{equation}
where $Q_m$ is the measured water ingress (or oil outflow) rate, $Q$ is the true water ingress (or oil outflow) rate and $\epsilon_q$ denotes the error in the measurement, which is modelled by a lognormal distribution with median one and coefficient of variation $\delta_{\epsilon_q}=10\%$, if the quality of measurement is good (e.g., when sensors are installed), and $\delta_{\epsilon_q}=30\%$ if the quality of measurement is low. If the width opening extends to multiple tanks, the measured flow rate $Q_m$ is equal to the sum of the individual measured flow rates for each tank. The corresponding BN model is presented in Figure \ref{fig:Water ingress and oil spill}.

\subsection{Hydrostatics and bathymetric module}
Hydrostatics and stability data of the damaged ship (at the final floating position) in combination with bathymetric (water depth) data from the incident site can provide insights on the estimation of vertical damage extent $D_v$ and the transverse location of the center of the damage $Y_D$. 

\subsubsection{Hydrostatics and stability data}
\begin{figure}[h]
    \centering
    \includegraphics[width=0.75\linewidth]{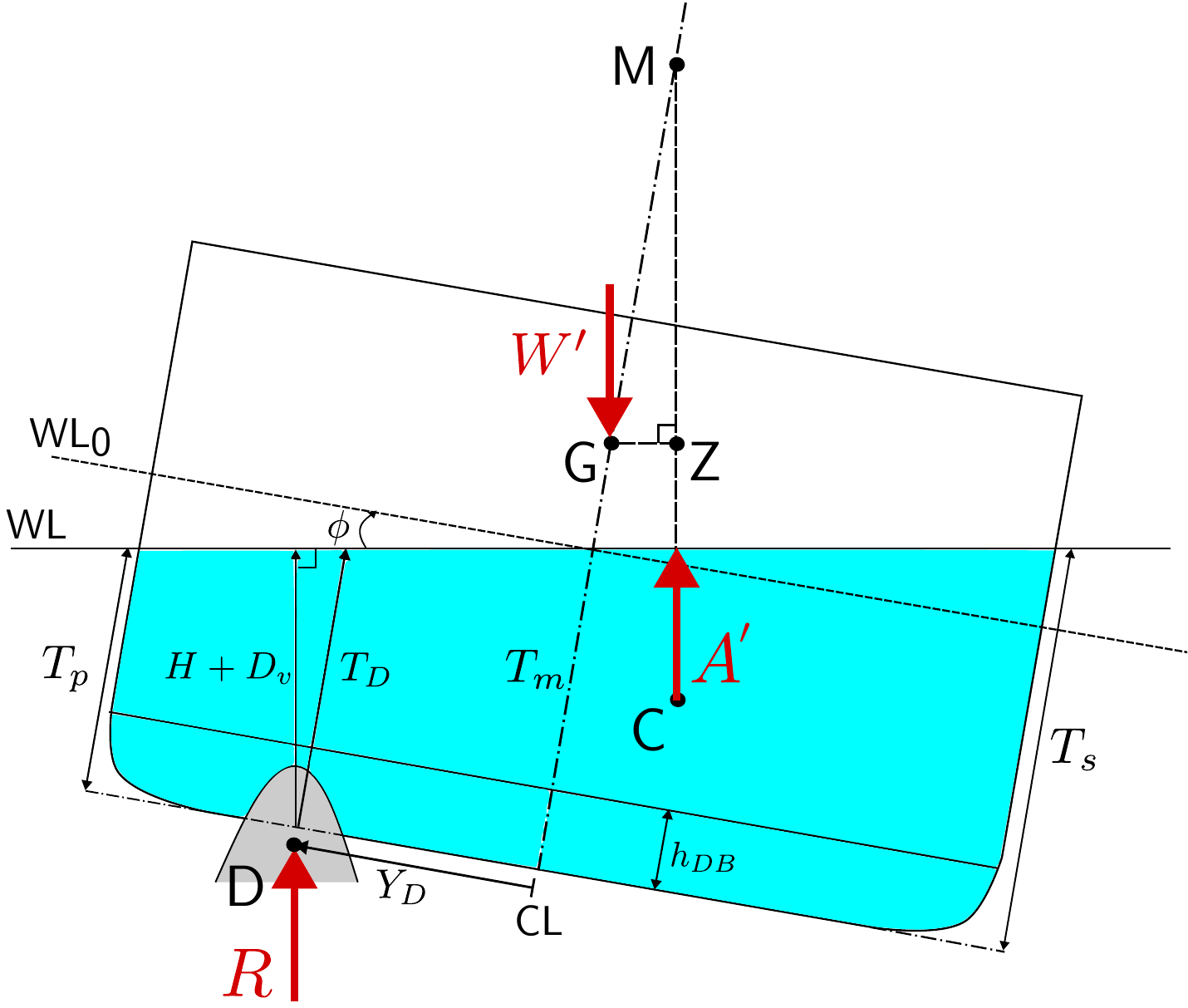}
    \caption{Ship in static equilibrium after stranding. The weight (mass) of the ship $W'$ is supported by the buoyancy $A'$ and the ground reaction $R$. The center of gravity G after damage is generally different than the one in the intact condition; here we assume symmetric flooding, so G remains on the centerline CL.}
    \label{fig:Drafts}
\end{figure}

The BN model is developed for the case in which the ship is stranded on a rock (or pinnacle). In this condition, the ship balances under the action of displacement $W'$ (or mass $M'$) -- after considering the inflow and outflow masses of water ingress and oil spill--, ground reaction $R$ and buoyancy $A'=W'-R$, as shown in Figure \ref{fig:Drafts}. The moment equilibrium condition around the centerline plane requires the heeling moment caused by the ground reaction and the righting moment of the ship to be equal, that is:

\begin{equation} \label{Moments}
     R\cdot Y_D  = (W' - R) \cdot GM \cdot \tan{\phi}
\end{equation}
where $GM$ is the transverse metacentric height in the damaged condition, $Y_D$ is the distance from the centerline to the center of the damage (positive to port side), which is assumed to coincide with the position of ground reaction, and $\phi$ is the heeling angle (positive if the rock is on the port side) given as:

\begin{equation} \label{phi}
    \tan{\phi} = \frac{T_s - T_p}{B} 
\end{equation}
where $T_p$ is the port draft and $T_s$ is the starboard draft at the critical damage location along the length of the ship. In the BN, $T_s$ is introduced as a child of $\phi$ and $T_p$, by solving the deterministic relation of Eq. (16) for $T_s$. An uncertainty on the observed drafts is expected due to the imprecise identification of critical cross-section and due to wave action on the hull. Thus, the following relationships are considered:

\begin{align} \label{T0}
    T_{p,m} &= T_p + \epsilon_{tp} \\
    T_{s,m} &= T_s + \epsilon_{ts}
\end{align}
where $T_{p(s),m}$ is the measured draft port (resp. starboard), $T_{p(s)}$ is the true draft port (resp. starboard) and $\epsilon_{tp(s)}$ denotes the uncertainty on the measured values which is modelled by a normal distribution with zero mean and standard deviation $\sigma_{\epsilon_{tp(s)}}$. Here we assume $\sigma_{\epsilon_{tp}}=\sigma_{\epsilon_{ts}}=0.25$ m.

The ground reaction $R$ can be estimated using different methods. The simplest and most robust approach -- particularly when the exact location of the ground reaction is unknown -- is the displacement method \citep{Dean2016}. In this method, the reaction force is determined as the difference between the ship's displacement before and after stranding. Further details can be found in \citet{Dean2016}. The calculation of the ground reaction is performed outside the BN and the result is entered as evidence in the BN. Since all the methods provide approximate results, an uncertainty in the calculated value of ground reaction $R_c$ is assumed of the following form:

\begin{equation} \label{R}
    R_c = R \times \epsilon_r
\end{equation}
where $R$ is the true ground reaction force and $\epsilon_r$ denotes the uncertainty in the calculated value, which is modelled by a lognormal distribution with median equal to one and coefficient of variation $\delta_{\epsilon_r}=10\%$. Note that ground reaction varies in time due to tide effects and changes in the weight of the ship. The assigned value of $\delta_{\epsilon_r}$ also considers these time-dependent effects.

The metacentric height after grounding $GM$ can also be calculated outside the BN by taking into account the new displacement of the ship and the updated floating position. Ship damage stability calculations using dedicated tools (e.g., NAPA commercial software) can be performed. In general, it is expected that $GM$ is smaller than that of the intact condition $GM_0$ \citep{PNA1988}.

\begin{figure}
  \centering
    \begin{tikzpicture}[shorten >= 1pt, auto, thick]
        
    \node[latent] (Heel) {$\phi$};
    \node[latent, above=1.0 of Heel] (Mass new) {$M'$};
    \node[latent, above=1.0 of Heel, xshift=1.5cm] (Ground force) {$R$};
    \node[latent, fill=green!20, below=0.5 of Ground force, xshift=1.0cm] (Ground force rep) {$R_c$};
    \node[latent, fill=yellow!50, above=1.0 of Heel, xshift=-1.5cm] (Damage position) {$Y_D$};
    \node[latent, below=1.0 of Heel] (Draft stbd) {$T_s$};
    \node[latent, fill=green!20, below=0.5 of Draft stbd, xshift=1.0cm] (Draft stbd rep) {$T_{s,m}$};
    \node[latent, left=2.0 of Draft stbd] (Draft port) {$T_p$};
    \node[latent, fill=green!20, below=0.5 of Draft port, xshift=-1.0cm] (Draft port rep) {$T_{p,m}$};    
    \node[latent, below=5.0 of Damage position] (Draft damage) {$T_D$};
    \node[latent, fill=yellow!50, below=1.0 of Draft damage, xshift=1.0cm] (Penetration) {$D_v$};
    \node[latent, right=1.0 of Draft damage] (Water depth) {$H$};
    \node[latent, fill=green!20, below=0.5 of Water depth, xshift=1.0cm] (Water depth rep) {$H_r$};
    
    \edge {Ground force} {Ground force rep}
    \edge {Ground force} {Heel}
    \edge {Damage position} {Heel}
    \edge {Damage position} {Heel}
    \edge {Damage position} {Draft damage}
    \edge {Draft port} {Draft damage}
    \edge {Draft stbd} {Draft damage}
    \edge {Heel} {Draft stbd}
    \edge {Draft port} {Draft stbd}
    \edge {Draft port} {Draft port rep}
    \edge {Draft stbd} {Draft stbd rep}
    \edge {Draft damage} {Penetration}
    \edge {Water depth} {Penetration}
    \edge {Water depth} {Water depth rep}
    \edge {Mass new} {Heel}

    \end{tikzpicture}
\caption{BN model based on hydrostatics, damage stability and bathymetric data.}
\label{fig:BN Yd}   
\end{figure}
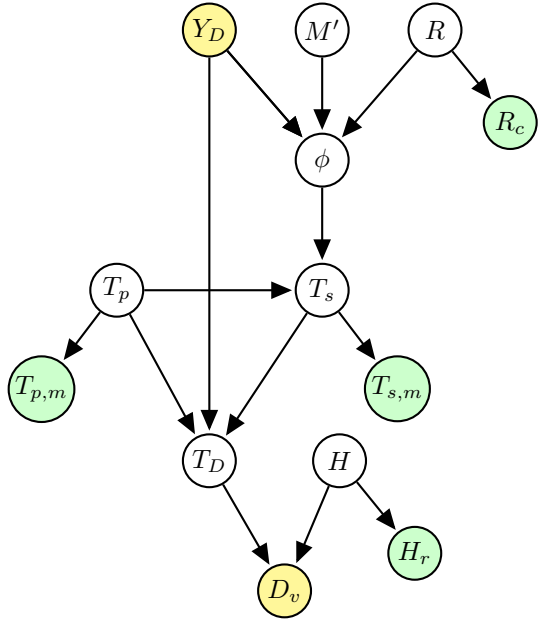

Having determined the parameter $Y_D$ through Eq. \ref{Moments}, the draft $T_D$ on the pinnacle location can be calculated from the geometry of Figure \ref{fig:Drafts} as:

\begin{equation} \label{TD}
    T_D = T_m - Y_D\tan{\phi}   
\end{equation}
where $T_m=(T_p+T_s)/2$ is the mean draft at the cross-section. Finally, the penetration depth $D_v$ can be inferred conditional on the water depth $H$ (i.e., the vertical distance from the sea surface to the rock tip) and the draft $T_D$ as:

\begin{equation} \label{Dv}
    D_v = T_D \cdot {\cos{\phi}} - H \approx T_D - H
\end{equation}
where we consider that the heeling angel $\phi$ is generally small. The corresponding BN is presented in Figure \ref{fig:BN Yd}.

\subsubsection{Water depth estimation}
The water depth can be estimated from onboard electronic navigational charts or bathymetric maps, such as the General Bathymetric Chart of the Oceans (GEBCO). In turn, the location of the ship can be obtained from the Automatic Identification System (AIS). An uncertainty in the reported water depth is expected due to imprecise grounding location, imprecise bathymetric data, and tidal effects. We model this uncertainty through the following relationship:

\begin{equation} \label{Hr}
    H_r = H + \epsilon_h 
\end{equation}
where $H_r$ is the reported water depth, $H$ is the true water depth (from the sea surface to the rock tip) and $\epsilon_h$ denotes the error in the reported value, which is modelled by a normal distribution with zero mean and standard deviation $\sigma_{\epsilon_h}$. An upper value of the error due to imprecise bathymetric data is considered to be equal to 0.25 m \citep{IHO2020}, due to imprecise grounding location equal to 1.0 m, and and due to tidal effects equal to 0.25 m (tidal effects on the specific location should be accounted for in real cases). Assuming that these upper values correspond to the 95\% confidence interval, it is $\sigma_{\epsilon_h}=0.75$ m. 

An upper bound of $H$ can be set since the water depth cannot be greater than the maximum draft of the ship before grounding $T_{0,max}$. A uniform distribution in the range 0 to $T_{0,max}$ is set as a prior for $H$. If bathymetric data are available for the area, a more informative prior can be used.

\subsection{Inspection module}
When applicable, underwater inspections can offer valuable insights into the extent and location of damage. The findings are typically presented as sketch reports, which inherently carry some degree of uncertainty. This uncertainty is further affected by site conditions, such as poor visibility, strong currents, or a moving hull \citep{ABS2012}. 

\begin{figure}[h]
  \centering
    \begin{tikzpicture}[shorten >= 1pt, auto, thick]
        
    \node[latent, fill=yellow!50] (Damage t) {$D_t$};
    \node[latent, fill=yellow!50, xshift=+1cm] (Damage v) {$D_v$};
    \node[latent, fill=yellow!50, xshift=+2cm] (Damage loc) {$Y_D$};
    \node[latent, fill=green!20, below=1.0 of Damage t] (Divers 1) {$Z_t$};
    \node[latent, fill=green!20, below=1.0 of Damage v] (Divers 2) {$Z_v$};
    \node[latent, fill=green!20, below=1.0 of Damage loc] (Divers 3) {$Z_y$};
    \node[latent, fill=green!20, below=1.0 of Divers 2] (Visibility) {$Vis$};

    \edge {Damage t} {Divers 1}
    \edge {Damage v} {Divers 2}
    \edge {Damage loc} {Divers 3}
    \edge {Visibility} {Divers 1}
    \edge {Visibility} {Divers 2}
    \edge {Visibility} {Divers 3}

    \end{tikzpicture}
\caption{BN for the estimation of damage extent and location based on underwater inspection $\mathbf{Z}=[Z_t,Z_v,Z_y]^T$.}
\label{fig:BN divers}   
\end{figure}
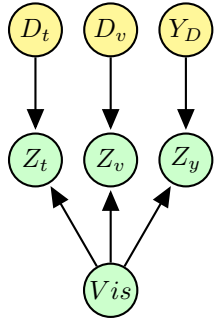

In the present work, we assume that the inspection outcome $\mathbf{Z}=[Z_t,Z_v,Z_y]^T$ is subject to a common influencing factor, the visibility $Vis$. The visibility node has two states, i.e., poor and good visibility. A-priori, each state has equal probabilities. The inspection outcome for each node of interest is defined conditional on the visibility node and the true damage state. The BN model of Figure \ref{fig:BN divers} is constructed.

For the damage width and vertical extent, we assume that the outcome of the inspection is expressed as:

\begin{equation} \label{Zt}
    Z_{t(v)} = D_{t(v)} \times \epsilon_d
\end{equation}
where $Z_{t(v)}$ is the observed value of transverse (resp. vertical) damage extent, $D_{t(v)}$ is the true, but unknown, damage transverse (resp. vertical) extent and $\epsilon_d$ is the observation error that follows a lognormal distribution with median equal to one and coefficient of variation $\delta_{\epsilon_d}$. Depending on the visibility node, we assume $\delta_{\epsilon_d}=10\%$ for good visibility and $\delta_{\epsilon_d}=30\%$ for poor visibility.

For the damage width location, we assume that the outcome of the underwater inspection is expressed using the following relationship:

\begin{equation} \label{Zy}
    Z_y = Y_D + \epsilon_y
\end{equation}
where $Z_y$ is the observed value of width center location, $Y_D$ is the true width center location and $\epsilon_y$ is the observation error which is modelled by a normal distribution with zero mean and standard deviation $\sigma_{\epsilon_y}=1$ m, if the visibility is good and $\sigma_{\epsilon_y}=2$ m, if the visibility is poor.

For the modelling of $\epsilon_d$ and $\epsilon_y$ above, we consider that the diver's estimate is unbiased. In case where a biased estimate is provided by the inspector (usually an overestimation), the mean value should be shifted to values larger than zero.

\subsection{Full Bayesian Network}     \label{Full BN}
\begin{figure} []
    \centering
    \includegraphics[width=1.0\linewidth]{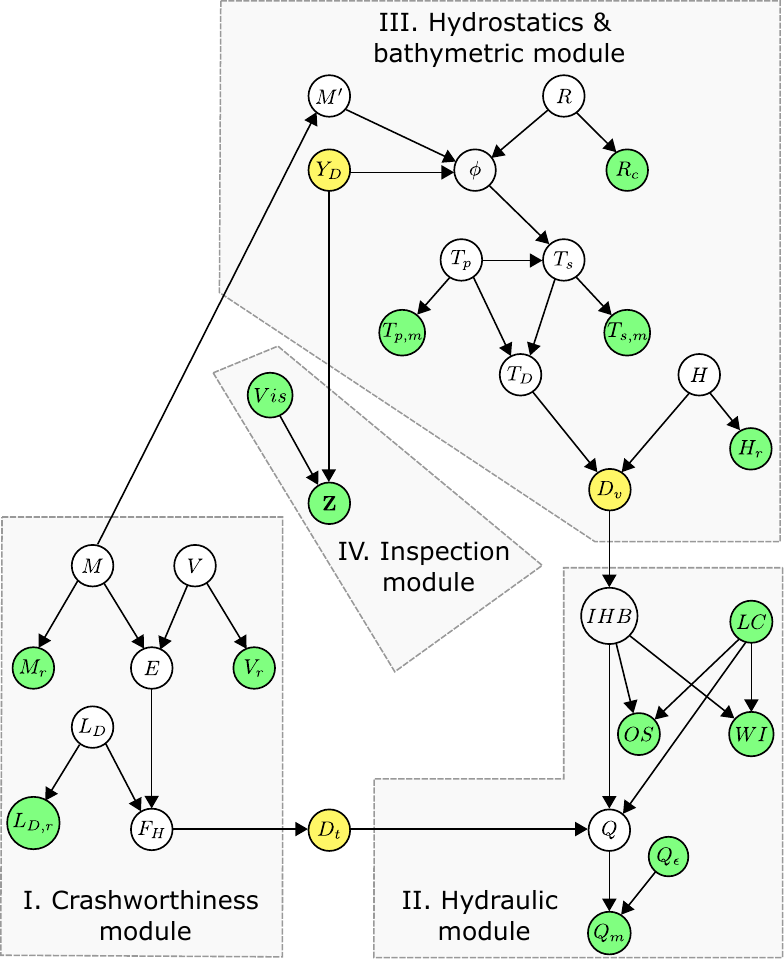}
    \caption{Full BN model for the 2D damage assessment problem. The nodes of interest are $D_t$, $D_v$ and $Y_D$.}
    \label{fig:BN full}   
\end{figure}

The complete BN of the problem is illustrated in Figure \ref{fig:BN full}. The complete set of variables determining the BN is listed in Table \ref{table:Summary RVs}. 

\begin{table}[]
\centering
\caption{Summary of random variables used in the BN of Figure \ref{fig:BN full}.}
\label{table:Summary RVs}
\begin{tabular}{lcc}
\hline
Variable                              & \multicolumn{1}{l}{Symbol} & \multicolumn{1}{l}{Units} \\ \hline
Damage length (reported)                                      & $L_D(L_{D,r})$                     & m                         \\
Displacement in damaged condition                                      & $M'$                        & t                         \\
Displacement initial (reported)                                      & $M(M_r)$                        & t                         \\
Draft aground port side (measured)                           & $T_p(T_{p,m})$                      & m                         \\
Draft aground starboard side (measured)                     & $T_s(T_{s,m})$                      & m                         \\
Draft aground at rock tip                                    & $T_D$                      & m                         \\
Flow rate (measured)                                        & $Q (Q_m)$                        & m\textsuperscript{3}/s                      \\
Flow rate measurement error                               & $Q_{\epsilon}$                        & -                         \\
Ground reaction (calculated)                                 & $R(R_c)$                       & t                         \\
Heeling angle                                             & $\phi$                      & degrees                        \\
Horizontal grounding force                                & $F_H$                      & MN                        \\
Impact speed (reported)                                   & $V(V_r)$                        & kn                        \\
Impact energy                                               & $E$                        & MJ                        \\
Inner hull breach                                          & $IHB$                       & -                         \\
Inspection quality                                      & $Vis$                      & -                         \\
Loading condition                                          & $LC$                       & -                         \\
Oil spill detection                                        & $OS$                       & -                         \\
Survey at transverse damage extent                         & $Z_t$                      & m                         \\
Survey at transverse damage location                        & $Z_y$                      & m                         \\
Survey at vertical damage extent                           & $Z_v$                      & m                         \\
Transverse damage extent                                    & $D_t$                      & m                         \\
Transverse damage location                                 & $Y_D$                      & m                         \\
Vertical damage extent                                     & $D_v$                      & m                         \\
Water depth (reported)                                     & $H(H_r)$                        & m                         \\
Water ingress detection area                               & $WI$                       & -                         \\          \hline
\end{tabular}
\end{table}

\section{Numerical investigations}     \label{Application}
We investigate the proposed BN model through two case studies. In Section \ref{Case study 1}, a validation study based on the crashworthiness module proposed in Section \ref{Crashworthiness model} is presented. In Section \ref{Case study 2}, two hypothetical grounding scenarios are examined for a double-hull VLCC tanker based on the full BN presented in Section \ref{Full BN}. The BN modeling and analysis have been implemented in the software package GeNIe \citep{GENIE}.

\subsection{Case study I: Single-hull tanker grounding off Singapore} \label{Case study 1}
\begin{figure}
    \centering
    \includegraphics[width=0.75\linewidth]{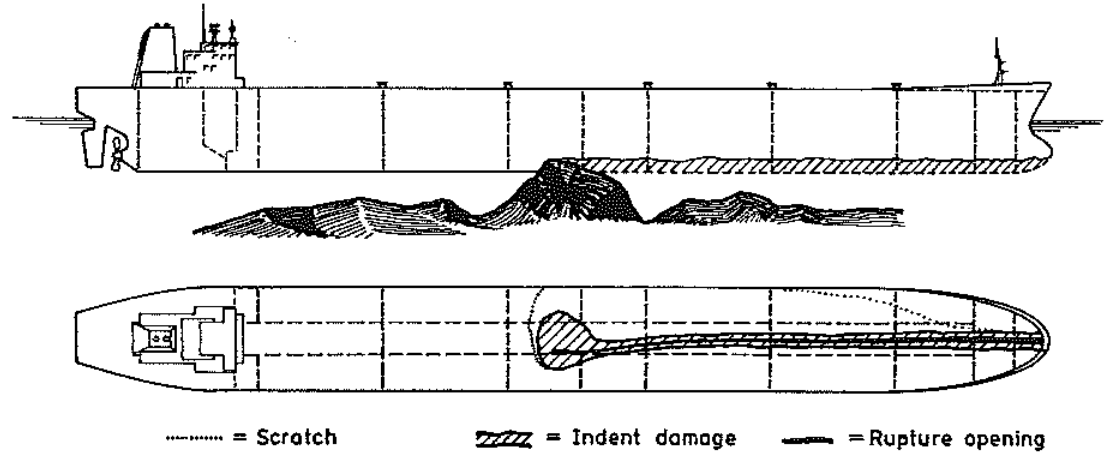}
    \caption{Grounding accident and resulting damage to a 273,000 t single-hull tanker \citep{Kuroiwa1996,Pedersen2000}.}
    \label{fig:single-hull grounding}
\end{figure}

In 1975, a newly built single-hull VLCC tanker ran aground off the coast of Singapore, resulting in an oil spill of more than 10,000 t. A schematic description of the damage is presented in Figure \ref{fig:single-hull grounding}. The reported damage length of the accident was about 180 m and the damaged width was between 6 to 10 m. The main particulars of the ship, along with other data reported during the accident, are listed in Table \ref{table:Ship data single-hull}. The bottom geometric and material characteristics are the following: $t_{eq}=62$ mm, $\sigma_0=300$ MPa and $\epsilon_f=0.41$.

\begin{table}[]
\centering
\caption{Principal particulars and reported values for the single-hull tanker grounding.}
\label{table:Ship data single-hull}
\begin{tabular}{lcc}
\hline
Main particulars        & Value   & Units \\ \hline
Length, $L$             & 304     & m     \\
Breadth, $B$            & 52.4    & m     \\
Depth, $D$              & 25.7    & m     \\
Draft (design), $T$     & 19.8    & m     \\
Service speed, $V_s$    & 15.0    & kn    \\
Impact speed, $V_r$     & 11.5    & kn    \\
Displacement, $M_r$     & 273,000 & t    \\ \hline
\end{tabular}
\end{table}

\subsubsection{BN modeling}
The BN structure of Figure \ref{fig:BN Energy} is deployed for validating the reported damage width. The prior distributions of the root nodes are listed in Table \ref{table:Priors A}. Evidence is provided in the nodes based on the reported data.

\begin{table}[]
\centering
\caption{Prior distribution models for single-hull tanker BN model.}
\label{table:Priors A}
\begin{tabular}{lcccc}
\hline
Variable    & Distribution     & Mean        & Standard dev.  & Bounds \\ \hline
$M$ [t]     & Uniform          & 250,000     & 28,868         & 200,000 to 300,000 \\
$V$ [kn]    & Beta             & 10.7        & 2.4            & 0 to 15 \\
$L_D$ [m]   & Empirical        & 67          & 65.2           & 0 to 304  \\ \hline
\end{tabular}
\end{table}

\subsubsection{Results}
\begin{figure}
    \centering
    \includegraphics[width=.75\linewidth]{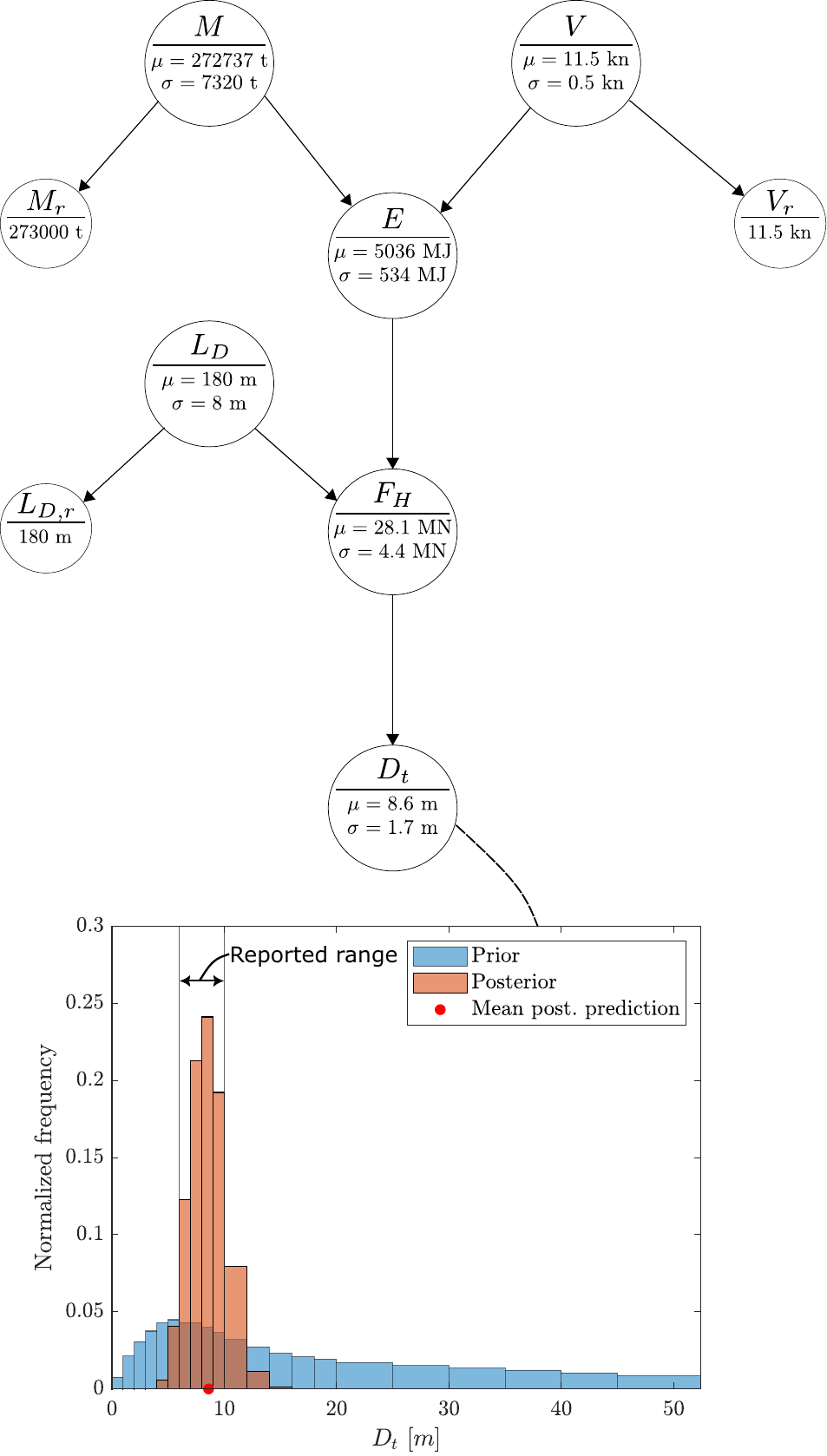}
    \caption{Validation of the crashworthiness module using real-world data from a grounded single-hull oil tanker. A very good agreement between the posterior distribution and the reported range from the incident is observed. The upper bound of the prior distribution has been truncated to the ship's breadth.}
    \label{fig:BN results validation}
\end{figure}

Figure \ref{fig:BN results validation} presents the results of the analysis. The conditional mean values and standard deviations are shown within the nodes after inserting the evidence. The posterior mean and standard deviation of $D_t$ are calculated as 8.6 m and 1.7 m. The same figure also displays the prior and the posterior normalized frequency histograms. A-priori, the damage width exhibits significant uncertainty over the plausible range from $0$ to $B$. After incorporating the evidence, the BN model prediction shows very good agreement with the true damage extent.

\subsection{Case study II. Hypothetical double-hull tanker grounding in the Gulf of Finland} \label{Case study 2}
In this case study, a double-hull VLCC tanker is assumed to run aground in the Gulf of Finland. The ship deviates from its planned route and strikes a rock while steaming straight ahead. The ship is in full load condition; all cargo tanks are filled to 98\% of their volume and the ballast tanks are empty. The principal particulars and the double bottom structural characteristics of the ship are listed in Table \ref{table:Ship data}. 

\begin{figure}
    \centering
    \includegraphics[width=1\linewidth]{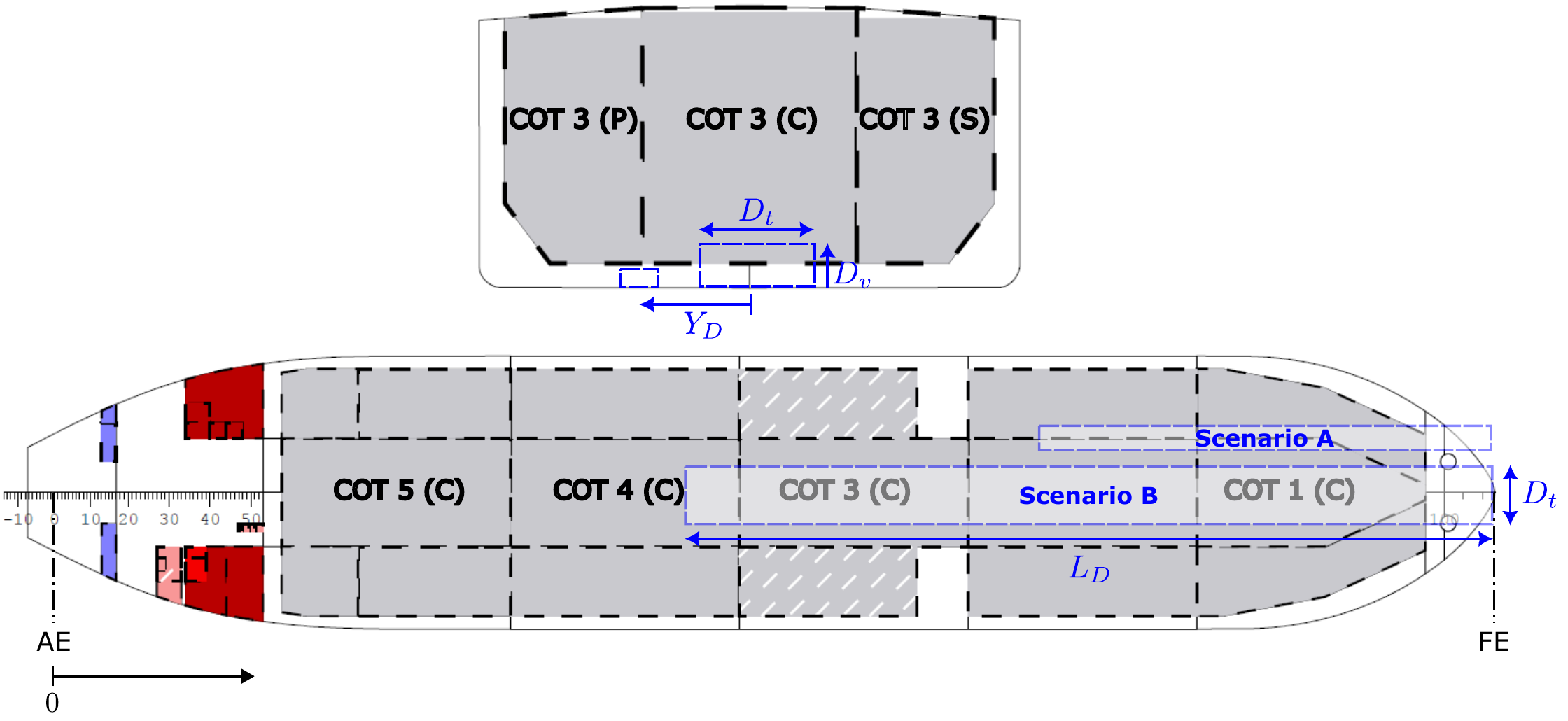}
    \caption{Schematics of the damage extent and location for the two grounding scenarios.}
    \label{fig:Damage scenarios}
\end{figure}

The purpose of this case study is to provide a verification of the proposed BN model. Two hypothetical grounding scenarios are examined. A schematic description of the two damage patterns is illustrated in Figure \ref{fig:Damage scenarios}. More details listed in Table \ref{table:Details grounding}. Two different site locations are assumed, with different water depths, resulting in different damage extents. The first scenario is less severe; only the outer bottom is breached. The second scenario is more severe, leading to inner bottom penetration and a subsequent oil spill. The ``true" values were assumed by introducing small perturbations to the input parameters of the modules. They also show slight deviations from the corresponding inspection outcomes.

\begin{table}[]
\centering
\caption{Principal particulars and double bottom structural characteristics of the double-hull VLCC tanker.}
\label{table:Ship data}
\begin{tabular}{lcc}
\hline
\multicolumn{1}{c}{Main particulars} & Value   & Units \\ \hline
Length, $L$                          & 316     & m     \\
Breadth, $B$                         & 60      & m     \\
Depth, $D$                           & 29.7    & m     \\
Draft (design), $T$                  & 19.2    & m     \\
Service speed, $V_s$                 & 15.6    & kn    \\
Metacentric height (initial), $GM_0$ & 7.84    & m      \\
Double bottom height, $h_{DB}$       & 2.7     & m     \\
Outer (Inner) bottom equivalent thickness, $t_{eq}$    & 45 (45)     & mm  \\
Outer (Inner) bottom flow stress, $\sigma_0$           & 427 (427)    & MPa  \\ 
Outer (Inner) bottom fracture strain, $\epsilon_f$     & 0.25 (0.25)   & -  \\ \hline
\end{tabular}
\end{table}

\begin{table}[]
\centering
\caption{Data of the two grounding scenarios for the double-hull VLCC tanker.}
\label{table:Details grounding}
\begin{tabular}{lcc}
\hline
Pre-grounded conditions             & Scenario A            & Scenario B \\ \hline
Loading condition                   & Loaded                & Loaded     \\
Displacement {[}t{]}                & 298,474               & 298,474    \\
Impact speed {[}kn{]}               & 4.5                   & 11.5       \\
Draft aft {[}m{]}                   & 20.2                  & 20.2       \\
Draft mid {[}m{]}                   & 19.2                  & 19.2       \\
Draft fwd {[}m{]}                   & 18.1                  & 18.1       \\
Water depth {[}m{]}                 & 16.0                  & 14.5       \\
                                    &                       &            \\
Grounding conditions                &             &  \\ \hline
First intact long. position {[}m{]} & 316 (FE)                   & 316 (FE) \\
Damaged length {[}m{]}              & 100                   & 180        \\
Tank damaged length {[}m{]}         & $\sim$ 35                  & 50.4       \\
Oil spill detection                 & no                    & yes        \\
Flooding rate {[}m\textsuperscript{3}/s{]}   & 1,350         & -          \\
Oil outflow rate {[}m\textsuperscript{3}/s{]} & -           & 1,400       \\
Draft port {[}m{]}                  & 17.2                  & 19.0       \\
Draft stbd {[}m{]}                  & 20.9                  & 18.0      \\
Heeling angle [deg]                 & 3.5                   & -1.0       \\
Mass in/out [t]                     & 31,291/0              & 0/5,000    \\
Ground reaction [t]                 & 9,636               & 17,520    \\
Metacentric height [m]              & 6.3                   & 5.9        \\
                                    &                       &            \\
Inspection outcome                  &             &  \\ \hline
Damage width extent                 & 3.5 m                 & 6.5 m      \\
Damage vertical extent              & 1.5 m                 & 3.5 m      \\
Damage width location               & 14.0 m (Port side)      & -1.5 m (Stbd side) \\
                                    &                       &            \\
True values (assumed)               &             &  \\ \hline
Damage width extent                 & 3.3 m                 & 6.0 m      \\
Damage vertical extent              & 1.0 m                 & 3.4 m      \\
Damage width location               & 14.5 m (Port side)      & -1 m (Stbd side) \\\hline
\end{tabular}
\end{table}

\begin{table}[]
\centering
\caption{Prior distribution models for double-hull tanker BN model.}
\label{table:Priors B}
\begin{tabular}{lcccc}
\hline
Variable    & Distribution     & Mean        & Standard dev.  & Bounds \\ \hline
$M$ [t]     & Uniform          & 240,000     & 63,509         & 130,000 to 350,000 \\
$V$ [kn]    & Beta             & 11.14       & 2.49           & 0 to 15.6  \\
$L_D$ [m]   & Empirical        & 69.5        & 67.8           & 0 to 316   \\
$R$ [m]     & Uniform          & -        & -           & 0 to 10,000   \\ 
$Y_D$ [m]   & Uniform          & 0           & 17.3           & -30 to 30   \\
$H$ [m]     & Uniform          & 10.1        & 5.8            & 0 to 20.2   \\
$T_p$ [m]   & Uniform          & 17          & 5.8            & 7 to 27     \\ \hline
\end{tabular}
\end{table}

\subsubsection{BN modeling}
The full BN model illustrated in Figure \ref{fig:BN full} is deployed. The prior distributions of the root nodes are presented in Table \ref{table:Priors B}. Evidence is provided in the nodes based on the data listed in Table \ref{table:Details grounding}. 

\subsubsection{Results}    \label{Results Case 2}
\begin{figure}
    \centering
    \begin{subfigure}{0.45\textwidth}
    \centering
    \includegraphics[width=\textwidth]{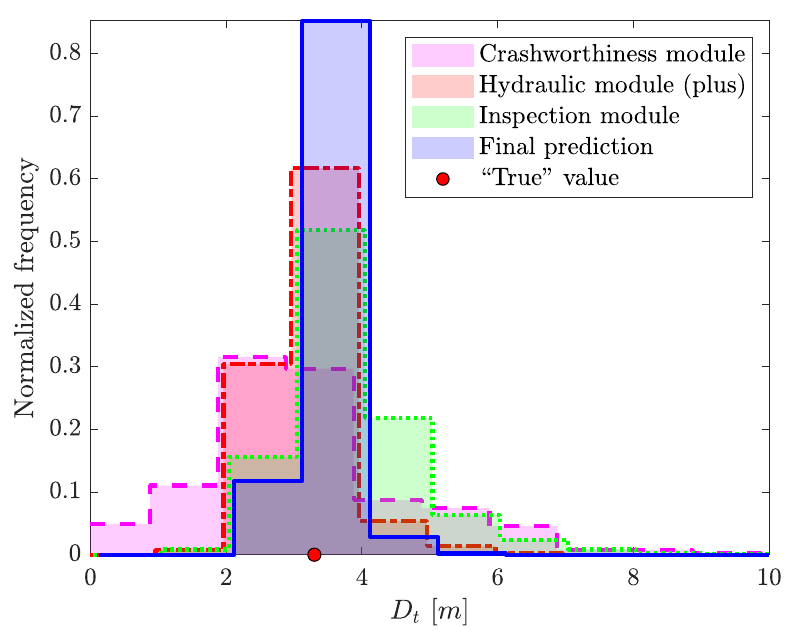}
    \caption{Damage width extent.}
    \label{fig:Results Scenario A Dt}
    \end{subfigure}   
    \hfill 
    \begin{subfigure}{0.45\textwidth}
    \centering
    \includegraphics[width=\textwidth]{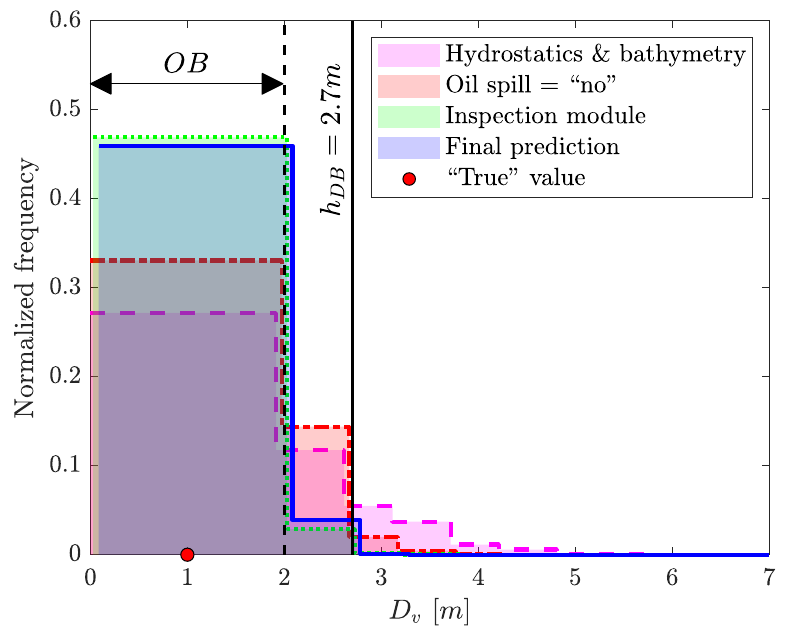}
    \caption{Damage vertical extent.}
    \label{fig:Results Scenario A Dv}
    \end{subfigure}
    \par
    \begin{subfigure}{0.45\textwidth}
    \centering
    \includegraphics[width=\textwidth]{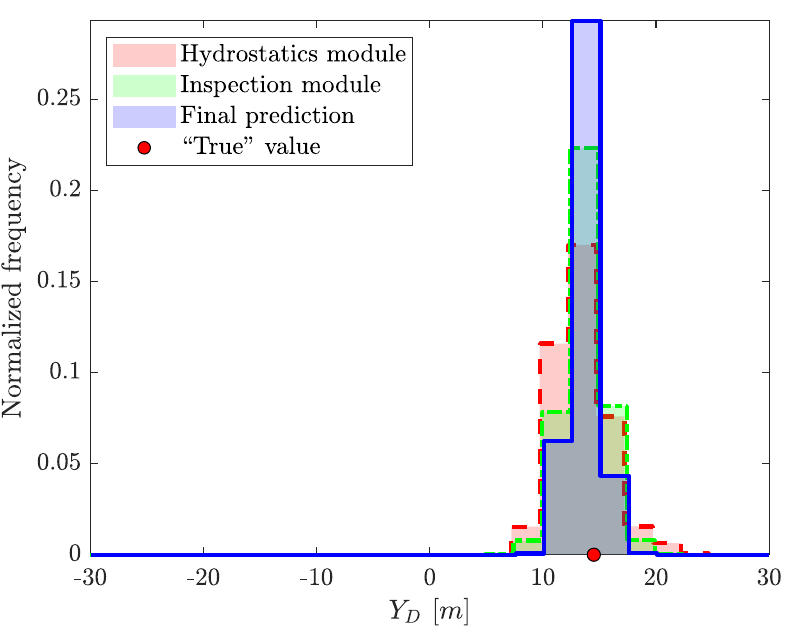}
    \caption{Transverse damage location.}
    \label{fig:Results Scenario A Yd}
    \end{subfigure}
    
    \caption{Posterior probability densities of damage extent and location (grounding scenario A).}
    \label{fig:Results Scenario A}
\end{figure}

\begin{figure}
    \centering
    \begin{subfigure}{0.45\textwidth}
    \centering
    \includegraphics[width=\textwidth]{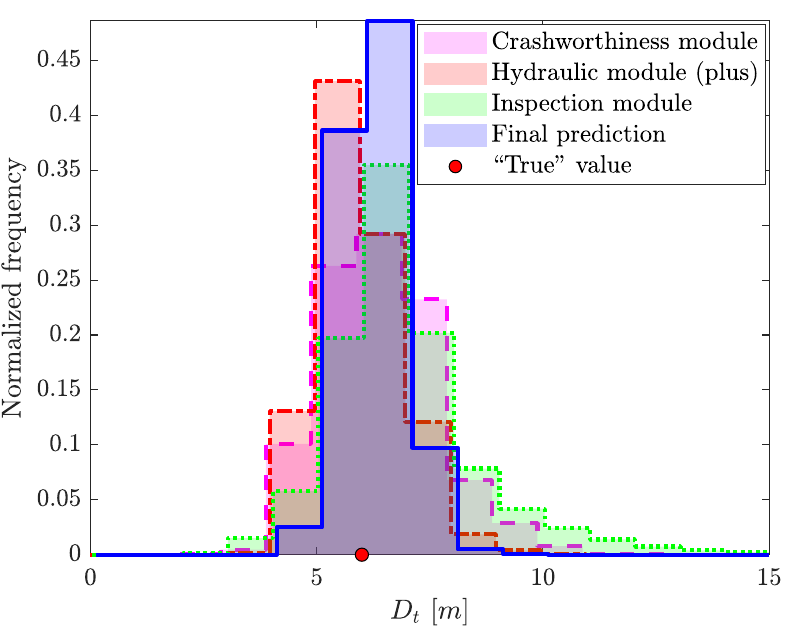}
    \caption{Damage width extent.}
    \label{fig:Results Scenario B Dt}
    \end{subfigure}   
    \hfill 
    \begin{subfigure}{0.45\textwidth}
    \centering
    \includegraphics[width=\textwidth]{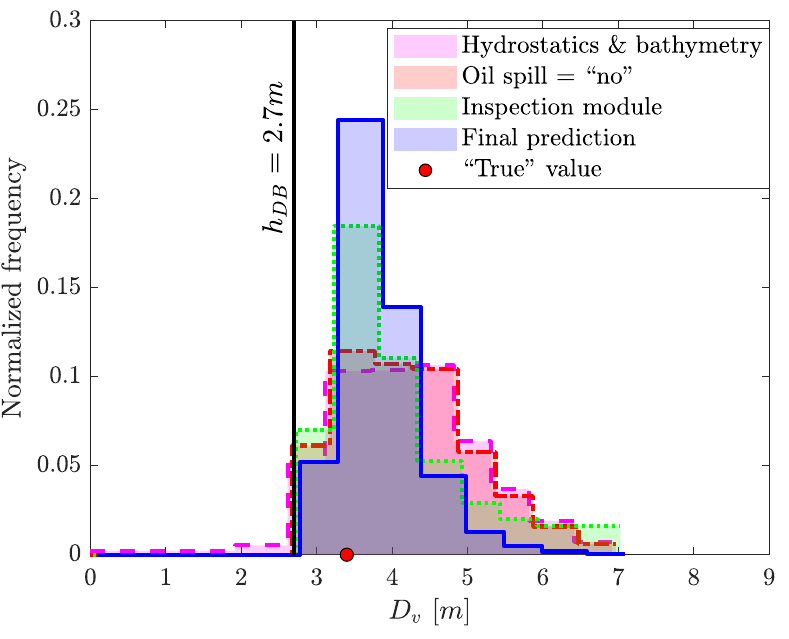}
    \caption{Damage vertical extent.}
    \label{fig:Results Scenario B Dv}
    \end{subfigure}
    \par
    \begin{subfigure}{0.45\textwidth}
    \centering
    \includegraphics[width=\textwidth]{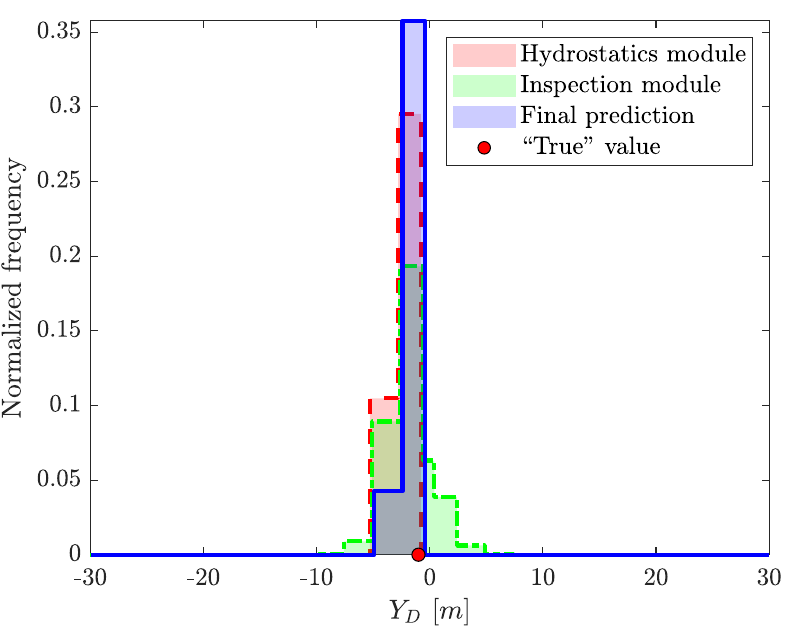}
    \caption{Transverse damage location.}
    \label{fig:Results Scenario B Yd}
    \end{subfigure}
    
    \caption{Posterior probability densities of damage extent and location (grounding scenario B).}
    \label{fig:Results Scenario B}
\end{figure}

The results for the two grounding scenarios are presented in Figures \ref{fig:Results Scenario A} and \ref{fig:Results Scenario B} as normalized frequency histograms. It is observed that as more evidence is gained, the posterior PDFs become narrower and peak around the underlying ``true" values for which the data of Table \ref{table:Details grounding} was created after a small perturbation on the input parameters. The posterior PDF of inspection module derives from the weighted average of good and poor visibility, corresponding to the (unlikely) case where the inspection conditions and quality are unknown. The effect of inspection quality as well as measurement quality of flow rate are examined in Section \ref{Sensitivity}. 

\paragraph{Damage scenario A}
The damage width extent $D_t$ prediction is depicted in Figure \ref{fig:Results Scenario A Dt}. The crashworthiness module can provide a first-cut rapid estimation of the mean value of $D_t$. However, the uncertainty in the estimate remains high. By additionally considering the hydraulic module, a significant decrease in the uncertainty of damage extent estimate is achieved. The prediction using the inspection module is similar to the one using both the crashworthiness and hydraulic modules, suggesting that the inspection can be replaced by the two modules. Nevertheless, performing an underwater inspection adds knowledge and the final prediction is even more accurate when combining information from all sources. 

The vertical damage extent $D_v$ prediction is presented in Figure \ref{fig:Results Scenario A Dv}. For all the cases presented, the prediction is very good, i.e., most of the probability density lies in the true (OB) state. The hydrostatics and bathymetric module provides a first good estimate of the damage extent that can be significantly improved by observing no oil spill, i.e., no inner hull breach. However, large plastic deformations and denting of structural elements are likely because the damage exceeds the double bottom height (see also Section \ref{Damage characterization}). In this case, it is recommended that the damaged parts be removed from the 2D structural analysis model of CSR. Finally, we observe that the inspection module has a high impact on the final prediction. 

The prediction of the transverse location of the damage $Y_D$ is illustrated in Figure \ref{fig:Results Scenario A Yd}. A uniform distribution over the breadth of the ship has been initially assumed. Using the hydrostatics module (i.e., moments equilibrium), a very good agreement between the predicted and the true value is observed. Incorporating information from the inspection will result in an even more accurate prediction.

\paragraph{Damage scenario B}
The results for the second damage scenario are presented in Figure \ref{fig:Results Scenario B}. Similar observations can be made for $D_t$ (see Figure \ref{fig:Results Scenario B Dt}) and $Y_d$ (see Figure \ref{fig:Results Scenario B Yd}) as in the first damage scenario. For the damage vertical extent $D_v$, the observation of oil spill shifts the probability density function to values larger than the double bottom height.

\begin{figure}
    \centering
    \begin{subfigure}{0.45\textwidth}
    \centering
    \includegraphics[width=\textwidth]{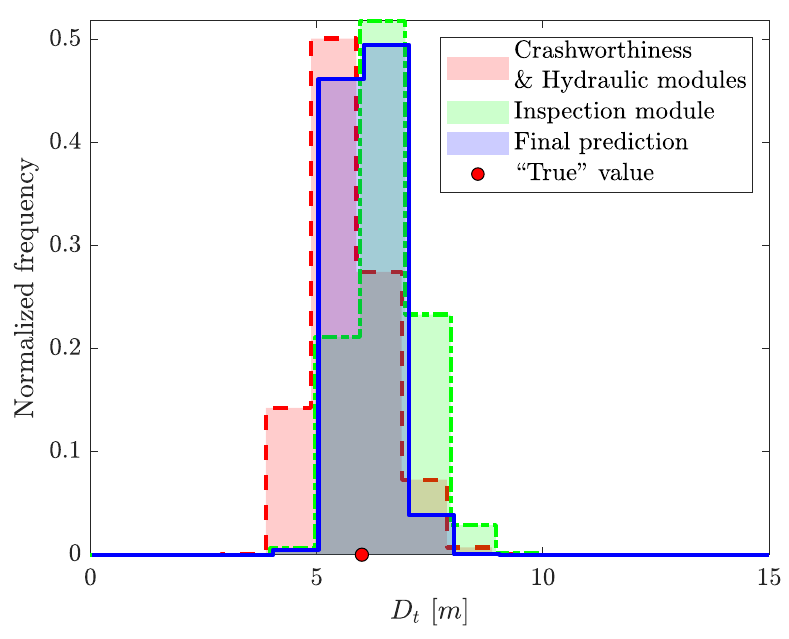}
    \caption{Flow measurement quality=``Good", Inspection quality=``Good".}
    \label{fig:Results Sens00}
    \end{subfigure}   
    \hfill 
    \begin{subfigure}{0.45\textwidth}
    \centering
    \includegraphics[width=\textwidth]{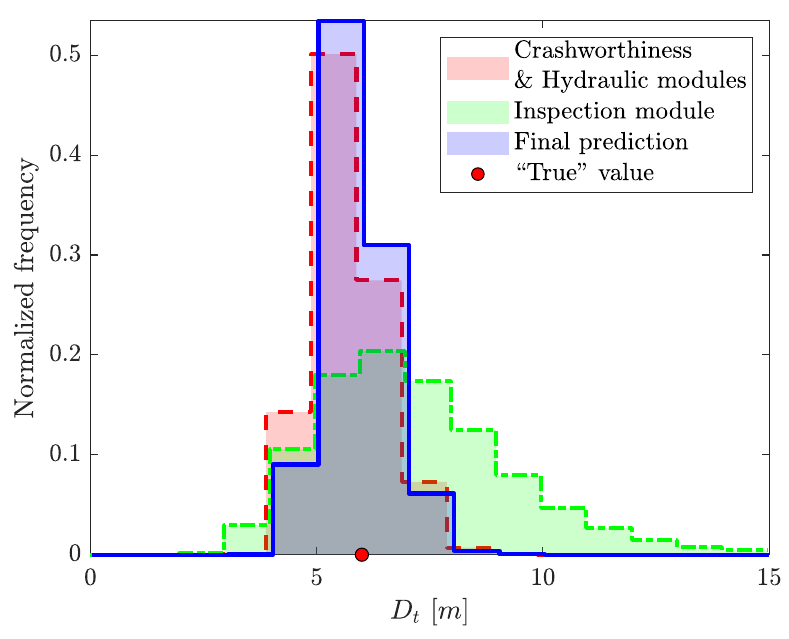}
    \caption{Flow measurement quality=``Good", Inspection quality=``Poor".}
    \label{fig:Results Sens01}
    \end{subfigure}
    \par
    \begin{subfigure}{0.45\textwidth}
    \centering
    \includegraphics[width=\textwidth]{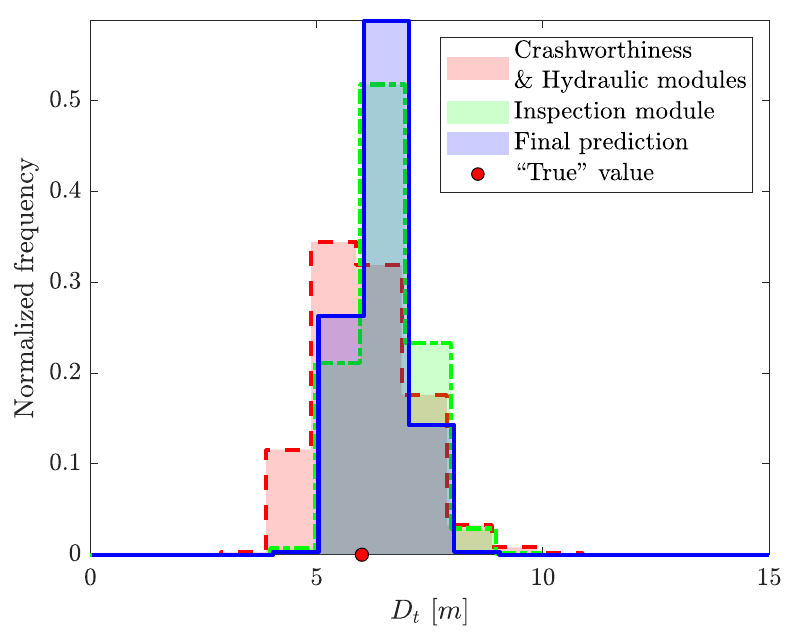}
    \caption{Flow measurement quality=``Poor", Inspection quality=``Good".}
    \label{fig:Results Sens10}
    \end{subfigure}
    \hfill
    \begin{subfigure}{0.45\textwidth}
    \centering
    \includegraphics[width=\textwidth]{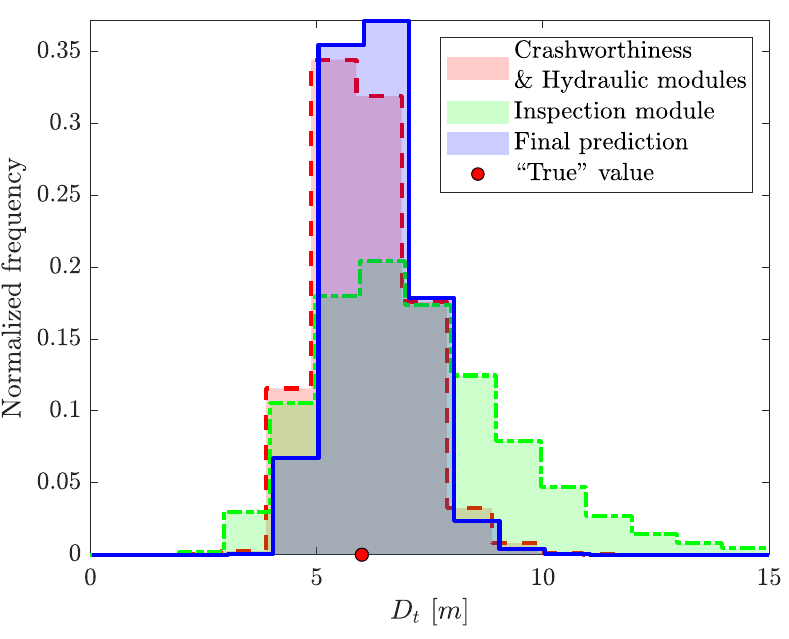}
    \caption{Flow measurement quality=``Poor", Inspection quality=``Poor".}
    \label{fig:Results Sens11}
    \end{subfigure}
    
    \caption{Sensitivity analysis results for damage scenario B and transverse damage extent $D_t$.}
    \label{fig:Results Sensitivity}
\end{figure}

\paragraph{Sensitivity study}       \label{Sensitivity}
The influence of inspection quality and measurement accuracy of the oil outflow rate on the damage width estimate in scenario B is examined. The results are presented in Figure \ref{fig:Results Sensitivity}. The findings indicate that the combination of the crashworthiness and hydraulic modules results in equally good predictions as those obtained using the inspection module alone. As illustrated in Figure \ref{fig:Results Sens11}, a low-quality inspection offers no useful information. Only a high-quality inspection improves the knowledge on the damage extent (see Figure \ref{fig:Results Sens00} and \ref{fig:Results Sens10}).

\section{Discussion}    \label{Discussion}
The extent of damage is one of the critical parameters in determining the residual strength of a ship following a grounding event. Here, we introduce a Bayesian network (BN) model for the probabilistic assessment of damage extent after such an event. The developed BN model offers several advantages over existing approaches: 

\begin{itemize}
    \item Explicit treatment of uncertainties: Unlike the master's assessment, which generally involves conservative estimates, the BN systematically accounts for uncertainties in various parameters affecting the problem and formalizes them using probabilistic analysis. Evidence from on-site observations and inspections is used to update prior information through Bayesian inference.
    \item Integration of multiple sources of information/modules: Crashworthiness models, hydraulic flow models, on-site observations, and inspection outcomes are combined in an intuitive and efficient way to provide a reliable estimation of the damage extent.
    \item Near-real-time assessment: The BN offers a rapid assessment of the damage extent, which is crucial in real-life scenarios.
\end{itemize}

The proposed BN model can be applied to real operational conditions, serving as an input for residual strength analysis models, such as Smith's method, or as a complement to R-D diagrams for rapid estimation of residual strength \citep{Paik2012,Kim2020,Li2022}. Given the importance of residual strength in decision-making, these findings can assist in determining whether a ship can safely continue its voyage to a repair yard \citep{Zhang2021}. The primary objective of this work is to develop a user-friendly digital assistance tool for rapid damage assessment after ship grounding. Initially, the BN model can supplement traditional empirical-based approaches, but with increased confidence and model improvements, it might replace the master's judgment and fully support decision-making of a classification society, a shipping company, or a salvage company. 

A limitation of the proposed BN model is that it does not account for time-varying effects. During the flooding process, for example, drafts and heeling angle change over time. In this study, we assume that the drafts are observed at the final equilibrium position. The extension of the model over time could be achieved through the adoption of a dynamic BN model \citep{Murphy2002}.

Another aspect that is not explicitly accounted for in this paper is the additional damage that can be caused by tides and waves when the ship rests on the sea floor. In particular, if a ship runs aground at high tide, it may be subjected to larger grounding damage and hull-girder loads as the tide drops and buoyancy is lost \citep{Nguyen2011}. Future refinements could incorporate these effects to improve damage predictions. Moreover, the developed BN is limited to single rock grounding. Future work could extend the model to identify multiple damage openings resulting from grounding on multiple rocks \citep{Zhu2002}. 

The proposed BN focuses on 2D damage assessment, aligning with the hull-girder residual strength evaluation using the Smith's model of CSR. However, it can easily be extended to three dimensions by introducing additional nodes and corresponding evidence to determine the location and extent of damage length. Insights from experienced ship masters can be utilized for that purpose. Finally, the model can be adapted to accommodate different ship types and more loading conditions. 

In this paper, we have demonstrated and partially validated the good performance of the model on two case studies. Further validation data should be collected from the application of the model in real applications. Nevertheless, we have trust in the model's practical usefulness despite the fact that only partial validation of the model is possible at present. The reason lies in the causal nature of the BN, which provides full transparency and enables users to critically question the predictions. We recommend that the model be initially used to complement existing approaches to assessing damage until more validation data is available.

\section{Conclusions}       \label{Coclusions}
In this paper, we introduce a BN model for rapid assessment of 2D damage extent and location following a hard grounding event. The model provides a probabilistic description of the damage by accounting for the joint influence of multiple uncertain variables, which are grouped into four interconnected modules: crashworthiness, hydraulic, hydrostatic and bathymetric, and inspection. The performance of the BN model is demonstrated through two case studies. In the first case study, we perform a validation of the BN model with a real grounding scenario. With the second case study, we present a verification of the BN model and we show that combining the crashworthiness and hydraulic modules with onboard observations and monitoring can even replace costly underwater inspections. The proposed model is ready for application in real operational conditions, but should still be complemented with traditional assessment methods until sufficient experience with the model is collected.

\section*{Data availability}         
The Bayesian network model developed in this study is implemented in GeNIe Modeler \cite{GENIE} and can be made available under reasonable request.

\section*{Acknowledgments}
This work was supported by the Alexander Von Humboldt Foundation. This support is gratefully acknowledged.

\bibliographystyle{elsarticle-harv} 

\end{document}